\documentclass[11pt,draftcls,onecolumn]{IEEEtran}

\usepackage{bm, bbm}
\usepackage{graphicx}
\usepackage[margin = 2cm]{geometry}
\usepackage{multirow}

\newcommand{\T}{^\top}
\usepackage{empheq}
\usepackage{mathrsfs}
\usepackage{framed}
\usepackage{graphicx}
\usepackage{cite}
\usepackage{amsmath}
\usepackage{amssymb}
\usepackage{dsfont}
\usepackage[mathcal]{euscript}
\usepackage{empheq}
\usepackage{graphics}
\usepackage{bm}
\usepackage{psfrag}
\usepackage{mathrsfs}
\usepackage{bbm}
\usepackage{color}
\usepackage{cancel}
\input{epsf}
\newtheorem{theorem}{Theorem}
\newtheorem{lemma}{Lemma}

\newtheorem{remark}{Remark}
\newtheorem{assumption}{Assumption}

\renewcommand{\P}{\mathbb{P}}
\newcommand{\E}{\mathbb{E}}

\newcommand{\beq}{\begin{equation}}
\newcommand{\eeq}{\end{equation}}
\newcommand{\beqa}{\begin{eqnarray}}
\newcommand{\eeqa}{\end{eqnarray}}
\newcommand{\dfz}{\triangleq}

\begin{document}
\title{Interplay between Topology and Social Learning over Weak Graphs}

\author{Vincenzo~Matta, Virginia~Bordignon, Augusto Santos, and Ali H. Sayed
\thanks{V.~Matta is with DIEM, University of Salerno,
via Giovanni Paolo II, I-84084, Fisciano (SA), Italy (e-mail: vmatta@unisa.it).

V. Bordignon and A.~H.~Sayed are with the \'Ecole Polytechnique F\'ed\'erale de Lausanne EPFL, School of Engineering, CH-1015 Lausanne, Switzerland (e-mails: \{virginia.bordignon, ali.sayed\}@epfl.ch).

A. Santos was with the Adaptive System Laboratory, EPFL, CH-1015 Lausanne, Switzerland (e-mail: augusto.pt@gmail.com). 

This work was supported in part by  grant 205121-184999 from the Swiss National Science Foundation (SNSF). 
An early version with partial results from this paper was presented at ICASSP~\cite{MattaSantosSayedICASSP2019}.
}
}

\maketitle

%%%%%%%%%%%%%%%%%%%%%%%%%%%%%%%%%%%%%%%%%%%%%%%
\begin{abstract}
We consider a distributed social learning problem, where a network of agents is interested in selecting one among a finite number of hypotheses. 
We focus on {\em weakly-connected} graphs where the network is partitioned into a {\em sending} part and a {\em receiving} part.
The data collected by the agents might be heterogeneous, meaning that different sub-networks might be governed by different statistical models. For example, some sub-networks might intentionally generate data from a fake hypothesis in order to influence other agents.
This work focuses on a two-step diffusion strategy where each agent: $i)$ updates individually its {\em belief function} using its {\em private} data; $ii)$ computes a new belief function by exponentiating a linear combination of the logarithmic beliefs of its neighbors.
We provide two main contributions. 
First, we examine {\em what} agents learn over weak graphs (social learning problem). 
We obtain closed-form analytical formulas for the limiting beliefs at the different agents, which allow us to examine in some detail the learning performance of each agent.
These formulas reveal how the agents' detection capability and the network topology interact to influence the asymptotic beliefs of the agents.
In particular, the formulas allow us to predict if and when a leader-follower behavior is possible, where some sending agents can control the mind of the receiving agents by forcing them to choose a particular hypothesis.
Second, we consider the dual or reverse learning problem that reveals {\em how} agents {\em learned}: given a stream of beliefs collected at a receiving agent, we would like to discover the global influence that any sending component exerts on this receiving agent (topology learning problem).
A remarkable and perhaps unexpected interplay between social and topology learning is observed: given $H$ hypotheses and $S$ sending components, topology learning can be feasible when $H\geq S$.
The latter being only a necessary condition, we then examine the feasibility of topology learning for two useful classes of problems.
%, namely, a {\em structured} Gaussian model and a randomized model with {\em diversity} among sending sub-networks.
%dual problem focus on two classes of problems to find sufficient conditions. 
%Under a {\em structured} Gaussian model where the sending components have a certain uniformity in their likelihood models, we show that topology learning is feasible only for $S=2$. 
%For higher $S$, 
The conducted analysis reveals that a critical element to enable faithful topology learning is a sufficient degree of {\em diversity} in the statistical models of the sending sub-networks.
\end{abstract}


\section{Introduction}
\IEEEPARstart{I}\,n a social learning problem, several agents linked through a network topology form their individual opinions about a phenomenon of interest (learning process) by observing the beliefs of their neighboring agents (social interaction)~\cite{ChamleyBook,AcemogluOzdaglar2011,Jadbabaie2013,PoorSPmag2013,ScaglioneSPmag2013,ScaglioneACM2013,NedicTAC2015}. One relevant paradigm for social learning is that of {\em weakly-connected} graphs, which are prevalent in social networks and have been studied in~\cite{YingSayed2016, Salami,Zhao}. Under this model, there are two categories of sub-networks: the {\em sending} sub-networks, which feed information to the {\em receiving} sub-networks without getting back information from them. 
This scenario is common over social networks. For example, a celebrity may have a large number of followers, whose individual opinions are not necessarily followed by the celebrity. 
Another example is that of media networks, which promote the emergence of opinions  by feeding data to users without paying attention to feedback from them.

This work addresses two fundamental challenges arising in the study of social learning problems. 
One challenge is to understand the fundamental mechanism and implications of specific social learning strategies on opinion formation. 
In particular, over weak graphs, it is critical to understand how the receiving agents are influenced by the sending sub-networks. 
It is not difficult to envisage that the network topology plays an important role in determining the opinion formation. This motivates the second problem, which can be regarded as a {\em dual} learning problem. 
Given observation of the receiving agents' behavior, we want to establish whether it is possible to learn the strength of connections (weighted topology) from the sending components to the receiving agents. 
This second question is useful in identifying the main sources for opinion formation over a network. 

\subsection{Related Work}
The goal of social learning is to let each individual agent create its own opinion (formally, choose one from among a finite set of hypotheses) through local consultation with its neighbors. 
One classical distinction for social learning models is Bayesian vs. non-Bayesian models. 
In the former category, the belief of each agent is obtained by computing posterior distributions through Bayes' rule~\cite{ChamleyBook,SmithSorensen2000,Acemoglu2010,AcemogluOzdaglar2011,AcemogluRevEc2011,PoorSPmag2013,Krishnamurthy2014}. In order to accomplish this task, each agent must have some detailed knowledge of the other agents' likelihoods, and some prior knowledge about the phenomenon of interest. 
In the latter category, this level of knowledge is not assumed and the agents implement suitable distributed algorithms to interact with their neighbors and to aggregate their beliefs into their own~\cite{DeGroot,Epstein2010,AcemogluGEB2010,Jad,Jad2,YingSayed2016, Salami,Zhao,NedicTAC2017,Javidi}. 
The present work considers non-Bayesian learning. 

There are many useful implementations for non-Bayesian learning. 
The implementations differ in terms of the rule the agents adopt to update their beliefs. 
One first distinction concerns the type of distributed strategy. 
In particular, we can distinguish between consensus~\cite{Jad,Jad2} or diffusion~\cite{Zhao,YingSayed2016,Salami,NedicTAC2017,Javidi} implementations. 
A second distinction regards the way the beliefs are combined. 
In~\cite{Jad,Zhao,Salami}, they are combined {\em linearly}, while a linear combination of the {\em logarithmic} beliefs is used in~\cite{Javidi,NedicTAC2017}. 
This latter form is motivated by the fact that, in many detection problems, the best detection statistic is given by the linear combination of log-likelihoods and not of likelihoods.
As a matter of fact, using the log-belief combination can help achieve an improved (i.e., faster) learning rate~\cite{Javidi}. 

Once a particular learning rule has been chosen, the behavior of opinion formation will depend heavily on the type of network where the information propagates. In this respect, the majority of prior works in the literature focus on {\em strongly-connected} networks. 
In these networks, there always exist paths  linking any two agents in both directions (which makes them connected) and, in addition, at least one agent has a self-loop and places some partial trusts in its own data (which makes them {\em strongly} connected).
Under a {\em homogeneous} setting where the underlying true hypothesis is the same for the entire network, it has been shown that over strongly-connected networks all agents are able to discover and agree on the true hypothesis. 
This result is available in~\cite{Zhao,Jad} for diffusion and consensus rules with linear belief combination, and in~\cite{Javidi} for the diffusion rule with linear log-belief combination. 
There are results available also for the {\em heterogeneous} setting, where different agents might have different data models, different likelihoods, and promote different opinions across the network. 
In particular, the diffusion rule with linear log-belief combination with a {\em doubly-stochastic} combination matrix is considered in~\cite{NedicTAC2017}, where it is shown that, over a strongly-connected network, all agents reach a common opinion, by minimizing cooperatively the sum of Kullback-Leibler (KL) divergences across the network. 

In contrast, the important case of {\em weakly-connected} networks has received limited attention and was addressed more recently in~\cite{YingSayed2016, Salami} by using the linear-belief-combination rule. 
Several interesting phenomena arise over weak graphs, which are absent from strongly-connected networks. 
The main difference in this work from~\cite{YingSayed2016, Salami} is that we now consider the following general setting: $i)$ {\em diffusion-type} algorithms with linear combination of {\em log-beliefs} (as opposed to the beliefs themselves), $ii)$ {\em heterogeneous} data and inference model, and $iii)$ {\em inverse} topology learning.

\subsection{Main Results}
This work leads to two main contributions. First, we characterize (Theorem~\ref{theor:limbelief}) the limiting (as learning time elapses) agents' belief through analytical formulas that depend in a transparent manner on inferential descriptors (Kullback-Leibler divergences) and network descriptors (limiting combination matrix power).
Some revealing behaviors can be deduced from these formulas. 
For example, we will be able to characterize a {\em mind-control} effect in terms of the interplay between the {\em detection capacity} of each agent and the {\em centrality} of the different agents. 
We will see that useful analogies arise with what has been observed in~\cite{Salami}. 
For example, some of the effects shown in~\cite{Salami} will now be shown to hold under greater generality since our formulas can be obtained by relaxing some assumptions used in~\cite{Salami}, in particular, the {\em all-truths-are-equal} assumption, and the {\em prevailing-signal} assumption.
However, we will observe also some remarkable distinctions with respect to~\cite{Salami}. For instance, we will observe that the individual belief of each receiving agent will necessarily collapse to (or concentrate at) a {\em single hypothesis} (which might be different across the agents). 
This is in contrast with~\cite{Salami}, where the beliefs of the receiving agents could end up assigning some probabilities to more than one hypothesis. The reason for this difference in behavior arises from the difference in the combination rule used in this work in comparison to~\cite{Salami}.

The second contribution concerns the {\em topology learning} problem. 
We are interested to learn the topology linking the receiving agents to the sending components. 
This question is interesting because it would then allow us to identify the main sources of information in a network and how they influence opinion formation. Nevertheless, the inverse topology problem is challenging because we will assume that we can only observe the beliefs evolving at the receiving agents. 
In particular, we will be able to recover some macroscopic topology information, in terms of the limiting weights that each receiving agent sees from each sending component. We call this a macroscopic information since these weights incorporate: $i)$ the global effect coming from {\em all} agents belonging to a sending component, and $ii)$ the effect of intermediate receiving agents linked to the receiving agent under consideration. The relevance in estimating these global weights relies on the fact that the limiting beliefs of the receiving agents depend solely on this aggregate information.

We will establish conditions under which topology inference becomes feasible. 
More specifically, given $H$ hypotheses and $S$ sending components, under the assumption of homogeneous statistical models within each sending component, we will ascertain that a necessary condition to achieve consistent topology learning is (Lemma~\ref{prop:necessary}):
\beq
\boxed{
H \geq S}
\label{eq:HgtS0}
\eeq
%namely, the number of hypotheses must be at least equal to the number of sending components. 
Once established a necessary condition, we will examine some useful models to see whether topology learning {\em can be} in fact achieved. 
We consider first a {\em structured} Gaussian model where: $i)$ the true underlying (Gaussian) distributions are distinct across the sending sub-networks; and $ii)$ the (Gaussian) likelihoods are equal across the sending sub-networks, and contain the true distributions. For this setting, we will show in Theorem~\ref{theor:TopologyGaussian} that topology learning is feasible only when $S=2$.
We then recognize that one fundamental element for topology learning is the {\em diversity} between the sending sub-networks. Adding this further element, we will establish in Theorem~\ref{theor:TopologyRandom} that the problem is feasible for {\em any} $S$ provided that~(\ref{eq:HgtS0}) holds, and even under more general (e.g., non-Gaussian) models.

In summary, we remark that there are two learning problems coexisting in our work: a social learning problem and a topology learning problem. 
The former is the primary, or direct inferential problem that the agents are deployed for.
The latter is the {\em dual}, or {\em reverse} problem, which is in fact based on observation of the output (the beliefs) of the primary learning problem. One useful conclusion stemming from our analysis is to reveal some unexpected interplay between these two coexisting learning problems --- see Sec.~\ref{sec:SLvsTL} further ahead.

{\em Notation}. 
We use bold font notation for random variables, normal fonts for their realizations. 
Capital letters are generally used for matrices, whereas small font letters for vectors or scalars. 
Given a matrix $M$, the symbol $M^{\dagger}$ denotes its Moore-Penrose pseudoinverse. 
The $L\times L$ identity matrix is denoted by $I_L$. 
Likewise, the $L\times 1$ vector of all ones is denoted by $\mathbbm{1}_L$.
The notation ``a.s.'' signifies ``almost-surely''.

\section{Background}

\subsection{Data Model and Inference}
We consider a network of $N$ agents that collect streaming data from the environment. 
Formally, the random variable $\bm{\xi}_{k,i}\in\mathcal{X}_k$ denotes the data at agent $k\in\{1,2,\ldots,N\}$ collected at time $i\in\mathbb{N}$. The data are assumed to be independent over time, whereas they can be dependent across agents (i.e., over space). 

We work under a {\em heterogeneous} setting. First, the space where the data are defined, $\mathcal{X}_k$, is allowed to vary across agents. For example, the data at different agents can be random vectors of different sizes. In the social learning context, this scenario is not that uncommon since different types of attributes can be observed by different users across the network. 
Second, the data $\bm{\xi}_{k,i}$ are assumed to be generated according to certain distributions $f_k(\xi)$, which are allowed to vary across the agents as well, namely, for $k=1,2,\ldots,N$:
\beq
\boxed{
\bm{\xi}_{k,i}\sim f_k~~~~~~\textnormal{[true distribution]}
}
\eeq 
This type of heterogeneity can arise in practice for several reasons, for example, some agents may intentionally inject fake data to let the other agents have a bias towards fake hypotheses.

Based on the available data $\{\bm{\xi}_{k,i}\}$, the network agents aim to solve an inferential problem that consists in choosing one state of nature from among a finite collection $\Theta=\{1,2,\ldots,H\}$, with $H$ denoting the number of possible hypotheses. 
To solve such inferential problem, the agents rely on a family of {\em likelihood functions}. 
Specifically, for $k=1,2,\ldots,N$, the likelihood function of agent $k$ is denoted by:
\beq
\boxed{
L_k(\xi|\theta),~~~ \xi\in\mathcal{X}_k, \, \theta\in\Theta~~~~\textnormal{[likelihoods]}
} 
\eeq
We will often write $L_k(\theta)$ instead of $L_k(\xi|\theta)$ for simplicity. 
In our treatment, we assume that the data can be modeled either as continuous or discrete random variables, with the same (continuous or discrete) nature across all agents.   
Accordingly, both the true distributions and the likelihoods will be either probability density or probability mass functions, depending on the considered type of random variables.

We remark that the considered model includes the possibility that the true distribution $f_k(\xi)$ is equal to a certain likelihood $L_k(\xi|\theta_0)$ (as assumed, e.g., in~\cite{Salami,Javidi}) and in this case $\theta_0$ can be considered as the true underlying hypothesis. More generally (e.g., in~\cite{NedicTAC2017}) the true distribution $f_k(\xi)$ is not equal to any of the likelihoods $L_k(\xi|\theta)$, which might happen when the agents have an approximate knowledge of the statistical models, and even the likelihood more similar to the true distribution can be a mismatched version thereof.
In order to quantify the dissimilarity between the true distribution and a certain likelihood, we will use the Kullback-Leibler (KL) divergence~\cite{CT}:
\beq
D[f_k || L_k(\theta)]
\dfz
\E_{f_k}\left[\log\frac{f_k(\bm{\xi}_k)}{L_k(\bm{\xi}_{k}|\theta)}\right],\quad k=1,2,\ldots,N.
\label{eq:KLdivfirst}
\eeq
We remark that we have written $\bm{\xi}_{k}$ in place of $\bm{\xi}_{k,i}$ to highlight identical distributions across time, and that the expectation is computed under the actual distribution of $\bm{\xi}_k$, i.e., under $f_k$. 
In the forthcoming treatment we use the following regularity condition~\cite{NedicTAC2017,Javidi}.
\begin{assumption}[Finiteness of KL Divergences]
\label{assum:divergences}
All the KL divergences in~(\ref{eq:KLdivfirst}) are well-posed, namely, for all $k=1,2,\ldots,N$, and all $\theta\in\Theta$ we have that: 
\beq
\boxed{
D[f_k || L_k(\theta)]<\infty}
\label{eq:finiteKLdiv}
\eeq~\hfill$\square$
\end{assumption}

\begin{remark}[Likelihood Support]
\label{rem:supports}
Assumption~\ref{assum:divergences} implies that the likelihood $L_k(\bm{\xi}_{k,i}|\theta)$ can be equal to zero only for an ensemble of realizations $\bm{\xi}_{k,i}$ having zero probability under the distribution $f_k$.~\hfill$\square$ 
%Indeed, were $L_k(\bm{\xi}_{k,i}|\theta)=0$ with nonzero probability, then the KL divergence $D[f_k||L_k(\theta)]$ would be infinite, and this would contradict Assumption~\ref{assum:divergences}.
\end{remark}

\subsection{Social Learning Algorithm}
Motivated by the diffusion strategy in~\cite{Salami,Zhao} for opinion formation over social networks, in this article we consider the useful variations studied in~\cite{Javidi,NedicTAC2017}. 
The learning procedure employs a two-step algorithm that can be described as follows.
For each admissible hypothesis $\theta\in\Theta$ at time $i$, each agent $k$ uses its own fresh {\em private} observation, $\bm{\xi}_{k,i}$, to compute the local likelihood $L_k(\bm{\xi}_{k,i}|\theta)$. Using this likelihood, agent $k$ updates its local belief, $\bm{\mu}_{k,i-1}(\theta)$, obtaining an intermediate belief $\bm{\psi}_{k,i}(\theta)$ through the following update rule:
\beq
\boxed{
\bm{\psi}_{k,i}(\theta)=
\displaystyle{
\frac
{\bm{\mu}_{k,i-1}(\theta)L_k(\bm{\xi}_{k,i} | \theta)}
{\displaystyle{\sum_{\theta^{\prime}\in\Theta} \bm{\mu}_{k,i-1}(\theta^{\prime})L_k(\bm{\xi}_{k,i} | \theta^{\prime})}}
}
}
\label{eq:private}
\eeq
Then, agent $k$ aggregates the intermediate beliefs received from its neighbors through the following combination rule (the division by the denominator term in~(\ref{eq:combine}) is meant to ensure that $\bm{\mu}_{k,i}(\theta)$ is a probability measure with its entries adding up to one):
\beq
\boxed{
\bm{\mu}_{k,i}(\theta)=
\displaystyle{
\frac{
\exp\left\{
\displaystyle{\sum_{\ell=1}^N a_{\ell k} \log \bm{\psi}_{\ell,i}(\theta)}
\right\}
}
{
\displaystyle{
\sum_{\theta^{\prime}\in\Theta} \exp\left\{
\sum_{\ell=1}^N a_{\ell k} \log \bm{\psi}_{\ell,i}(\theta^{\prime})
\right\}
}
}
}
}
\label{eq:combine}
\eeq
where $a_{\ell k}$ is the {\em nonnegative combination weight} that agent $k$ uses to scale the intermediate log-belief received from agent $\ell$. It is assumed that $a_{\ell k}$ is equal to zero if $k$ does not receive information from $\ell$, which means that agent $k$ can combine only intermediate beliefs received from its neighbors. 
When collected into a {\em combination matrix} $A$ (with $(\ell,k)$-entry equal to $a_{\ell k}$), these combination weights are assumed to obey the standard requirements that make $A$ a left-stochastic matrix, namely, we have that:
\beq
A^{\top}\mathbbm{1}_N=\mathbbm{1}_N\Leftrightarrow \sum_{\ell=1}^N a_{\ell k}=1,~~~\forall k=1,2,\ldots,N.
\eeq
The value of $\bm{\mu}_{k,i}(\theta)$ provides an estimate for the likelihood by agent $k$ at time $i$ that the true hypothesis value is $\theta$. 
We remark that, differently from~\cite{Salami}, the second step in~(\ref{eq:combine}) combines linearly the {\em logarithm} of the intermediate beliefs, $\log \bm{\psi}_{\ell,i}(\theta)$, in the neighborhood of agent $k$.
Exponentiation and normalization are used to construct the final belief. 

We invoke the following standard initial condition, motivated by the fact that, at time $i=0$, the agents have no elements to discard any hypothesis~\cite{NedicTAC2017,Javidi}. 
\begin{assumption}[Initial Beliefs]
\label{assum:initbel}
All agents start by assigning strictly positive probability mass to all hypotheses, namely, $\mu_{k,0}(\theta)> 0$ for all $\theta\in\Theta$ and $k=1,2,\ldots,N$.~\hfill$\square$
\end{assumption}
\begin{remark}[Strict Positiveness of Beliefs]
\label{rem:posbeliefs}
From~(\ref{eq:private}) we see that, if $\bm{\mu}_{k,i-1}(\theta)>0$, then $\bm{\psi}_{k,i}(\theta)>0$ since $L_k(\bm{\xi}_{k,i}|\theta)>0$ (but for zero-probability sets) --- see Remark~\ref{rem:supports}. 
Strict positiveness of $\bm{\mu}_{k,i}(\theta)$ now follows from~(\ref{eq:combine}) since the combination weights are nonnegative and convex.~\hfill$\square$
\end{remark}

\section{Weak Graphs}
\begin{figure} [t]
\begin{center}
\includegraphics[scale= 0.37]{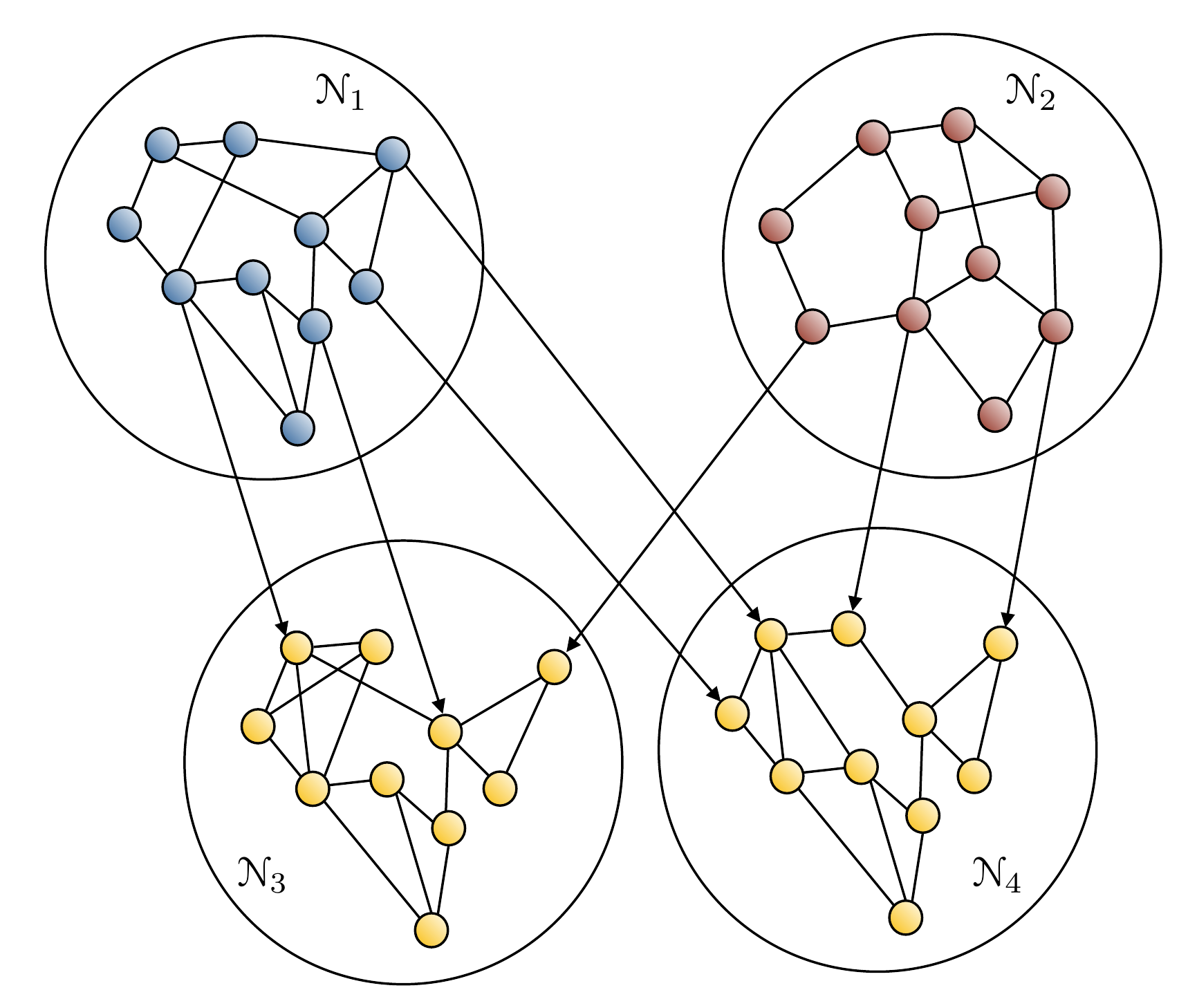}
\caption{One example of weakly-connected network, with sending sub-networks $\mathcal{N}_1$ and $\mathcal{N}_2$, and receiving sub-networks $\mathcal{N}_3$ and $\mathcal{N}_4$.}
\label{fig:weaklyconnparadigm}
\end{center}
\end{figure}

In this section, we consider the case of a weak graph (or weakly-connected network), which is defined as follows~\cite{YingSayed2016,Salami}. 
The overall network $\mathcal{N}=\{1,2,\ldots N\}$ is divided into $S+R$ disjoint components --- see Fig.~\ref{fig:weaklyconnparadigm}.
%\beq
%\mathcal{N}=\bigcup_{j=1}^{S+R} \mathcal{N}_{j}.
%\eeq 
The first $S$ sub-networks form the {\em sending} part $\mathcal{S}$, whereas the remaining $R$ sub-networks form the {\em receiving} part $\mathcal{R}$:
\beq
\mathcal{N}=\mathcal{S}\cup\mathcal{R},\qquad
\mathcal{S}\dfz\bigcup_{s=1}^S \mathcal{N}_{s},\qquad
\mathcal{R}\dfz\bigcup_{r=1}^{R} \mathcal{N}_{S+r}.
\eeq
We remark that $S$ and $R$ denote the number of sending and receiving {\em components}, respectively. They do {\em not} denote the cardinalities of $\mathcal{S}$ and $\mathcal{R}$, which are instead given by:
\beq
|\mathcal{S}|=\sum_{s=1}^S N_s,\qquad
|\mathcal{R}|=\sum_{r=1}^R N_{S+r},
\eeq
where the notation $N_j$ denotes the number of agents in sub-network $\mathcal{N}_j$.
Communication from a sending component to a receiving component is permitted, whereas communication in the reverse direction is forbidden. 
Communication between sending agents is possible, but a sending sub-network is identified by the following two conditions: $i)$ each sending sub-network is {\em strongly} connected~\cite{Sayed}; and $ii)$ agents belonging to different sending sub-networks do not communicate with each other. If some sending sub-networks communicate with each other, then they can be blended into a larger sending sub-network.
The $R$ receiving sub-networks are all individually assumed to be {\em connected} (but not necessarily {\em strongly} connected), with communication among the $R$ sub-networks being allowed. 
In particular, we assume that each receiving sub-network is connected to at least one agent in each sending sub-network.   

Without loss of generality, we assume that the network nodes across the $S+R$ components are listed in increasing order.
According to the above description, the combination matrix corresponding to a weakly-connected network admits the following convenient block decomposition~\cite{YingSayed2016,Salami}:
\beq
A=\left[\begin{array}{c|c}
   A_{\mathcal{S}} & A_{\mathcal{S}\mathcal{R}}\\
   \hline
   0 & A_{\mathcal{R}}
\end{array}\right]
\label{eq:Ablockstruct}
\eeq
where the matrix $A_{\mathcal{S}}={\sf blockdiag}\left\{A_{\mathcal{N}_1},A_{\mathcal{N}_2},\ldots,A_{\mathcal{N}_S}\right\}$
%\beq
%\label{eq:Asourceblock}
%\eeq 
contains the combination weights within the sending sub-networks, and has a block-diagonal form since communication between sending sub-networks is forbidden. 
The matrix block $A_{\mathcal{S}\mathcal{R}}$ contains the combination weights for the communication that takes place {\em from sending} agents {\em to receiving} agents. 
The left-bottom matrix block is zero since there are no direct links from the receiving agents to the sending ones. 
Finally, the matrix block $A_{\mathcal{R}}$ contains the combination weights ruling the communication among the receiving agents. 
Figure~\ref{fig:weaklyconnparadigm} offers a graphical illustration of the weakly-connected paradigm. 

It was shown in~\cite{YingSayed2016} that the limiting combination matrix power has the following structure:
\beq
A_{\infty}\dfz\lim_{i\rightarrow\infty}A^i=
\left[
\begin{array}{c|c}
E & E W \\
\hline
0 & 0
\end{array}
\right]=
\left[
\begin{array}{c|c}
E & \Omega \\
\hline
0 & 0
\end{array}
\right],
\label{eq:limatweak}
\eeq
where
\begin{equation}\label{eq:perron}
E={\sf blockdiag}\left\{p^{(1)}\mathbbm{1}^{\top}_{N_1},p^{(2)}\mathbbm{1}^{\top}_{N_2},\ldots,p^{(S)}\mathbbm{1}^{\top}_{N_S}\right\} 
\end{equation}
is a block diagonal matrix that stacks the $N_s\times 1$ Perron eigenvectors $p^{(s)}$ associated with the $s$-th sending sub-network\footnote{For $s=1,2,\ldots,S$, the Perron eigenvector of the sub-matrix $A_{\mathcal{N}_s}$ corresponding to the $s$-th sending sub-network is given by: 
\beq
A_{\mathcal{N}_s} p^{(s)}=p^{(s)},~~
\mathbbm{1}^{\top}_{N_s}p^{(s)}=1,~~p^{(s)}_{\ell}>0,~~\ell=1,2,\ldots,N_s.
\eeq
}, and where
\beq
W=A_{\mathcal{S}\mathcal{R}}\,(I_{|\mathcal{R}|} - A_{\mathcal{R}})^{-1}, \quad
\Omega=E W.
\label{eq:omegadef}
\eeq

We denote the entries of $\Omega$ by $[\omega_{\ell k}]$ and we keep indexing the columns of the $|\mathcal{S}|\times |\mathcal{R}|$ matrix $\Omega$ with an index 
\beq
k\in\left\{|\mathcal{S}|+1, |\mathcal{S}|+2,\ldots, |\mathcal{S}|+|\mathcal{R}|\right\}.
\label{eq:recindex}
\eeq
Since the limiting matrix power is left-stochastic and has a zero bottom block, the limiting weights $\omega_{\ell k}$ obey:
\beq
\sum_{\ell\in\mathcal{S}} \omega_{\ell k}=1,
\label{eq:convexomega}
\eeq
i.e., $\Omega$ is left-stochastic. 
From~(\ref{eq:omegadef}) we can also write: 
\beq
\Omega=E A_{\mathcal{S}\mathcal{R}}(I_{|\mathcal{R}|}+A_{\mathcal{R}}+A_{\mathcal{R}}^2+\dots),
\label{eq:Omegaexpand}
\eeq 
whence we see that $\omega_{\ell k}$ embodies the sum of influences over all paths from sending agent $\ell$ to receiving agent $k$.

\section{Limiting Beliefs of Receiving Agents}

Let us momentarily consider a single-agent scenario where agent $\ell$ operates alone. 
A natural way for agent $\ell$ to choose a hypothesis would be to choose the $\theta$ that gives the best match between a model $L_{\ell}(\theta)$ and the distribution of the observed data, $f_{\ell}$.
One measure of the match between $f_{\ell}$ and $L_{\ell}(\theta)$ is the KL divergence $D[f_{\ell}||L_{\ell}(\theta)]$. The smaller the value of this divergence is, the higher the match between the data and the model. 
For this reason, a strategy could be that of choosing the $\theta$ that minimizes the divergence $D[f_{\ell}||L_{\ell}(\theta)]$.

In the social learning context, this optimization problem turns into a {\em distributed} optimization problem.
In particular, under our social learning setting over weak graphs, we will show soon (Theorem~\ref{theor:limbelief}) that the log-belief diffusion strategy in~(\ref{eq:private})--(\ref{eq:combine}) will end up minimizing (without knowing the true distributions) the following {\em average} divergence at {\em receiving} agent $k\in\mathcal{R}$:
\beq
\boxed{
\mathscr{D}_k(\theta)\dfz \sum_{\ell\in\mathcal{S}} \omega_{\ell k} D[f_{\ell}||L_{\ell}(\theta)]
}
\label{eq:avDiv}
\eeq
which is a weighted combination, through the limiting combination weights $\{\omega_{\ell k}\}$, of the KL divergences of the {\em sending} agents reaching $k$.
The role of average divergence measures like the one in~(\ref{eq:avDiv}) already arose in the case of strongly-connected networks. For example, it was shown in~\cite{NedicTAC2017,Javidi} that with the log-belief diffusion strategy in~(\ref{eq:private})--(\ref{eq:combine}), each agent ends up minimizing the {\em same} weighted combination of divergences. Under classical identifiability conditions, such minimization leads each individual agent to discover the true underlying hypothesis~\cite{Javidi} or the best available approximation thereof~\cite{NedicTAC2017}. 
In our weak-graph setting, however, the effect of minimizing $\mathscr{D}_k(\theta)$ (which depends on the particular receiving agent $k$) will be less obvious. We already see from~(\ref{eq:avDiv}) that the average divergence combines {\em topological} attributes, encoded in the limiting combination weights, with {\em inferential attributes}, encoded in the local KL divergences. 
The interplay arising between the network topology and social learning will be critical in determining the choices of the receiving agents.

Throughout the work, we will invoke the following classical identifiability assumption that, as we will see in our examples, arises naturally in several models of interest.
\begin{assumption}[Unique Minimizer]
\label{assum:uniquemin}
For each $k=1,2,\ldots,N$, the function $\mathscr{D}_k(\theta)$ has a unique minimizer: 
\beq
\boxed{
\theta^{\star}_k\dfz\arg\!\min_{\theta\in\Theta}\mathscr{D}_k(\theta)
}
\label{eq:uniquemin}
\eeq~\hfill$\square$
\end{assumption}

We are now ready to characterize the limiting belief of the receiving agents. The following theorem is an extension to the case of weakly-connected graphs of similar theorems proved in~\cite{Javidi,NedicTAC2017} for the case of strongly-connected graphs.
\begin{theorem}[Belief Collapse at Receiving Agents]
\label{theor:limbelief}
Let $k\in\mathcal{R}$. Under Assumptions~\ref{assum:divergences}-\ref{assum:uniquemin} we have that:
\beq
\boxed{
\lim_{i\rightarrow\infty}\bm{\mu}_{k,i}(\theta^{\star}_k)\stackrel{\textnormal{a.s.}}{=}1
}
\label{eq:limbelief1st}
\eeq
where the symbol $\stackrel{\textnormal{a.s.}}{=}$ denotes that the pertinent limit exists almost-surely. 
Moreover, for all $\theta\neq \theta^{\star}_k$, the convergence of the belief to zero takes place at an exponential rate as:
\beq
\boxed{
\lim_{i\rightarrow\infty}\frac{\log\bm{\mu}_{k,i}(\theta)}{i}
\stackrel{\textnormal{a.s.}}{=}
\mathscr{D}_k(\theta^{\star}_k)-\mathscr{D}_k(\theta)
}
\label{eq:divergencescondition}
\eeq
\end{theorem}
\begin{IEEEproof}
The proof combines the techniques to establish the convergence of the social learning algorithm used, e.g., in~\cite{Javidi,NedicTAC2017} for strongly-connected graphs, with the convergence results of the combination matrix over weak graphs used in~\cite{Zhao,Salami}. 
The detailed steps are reported in Appendix~\ref{app:proofTheor1}.
\end{IEEEproof}

Several insightful conclusions arise from Theorem~\ref{theor:limbelief}. 

\begin{remark}[Collapse]
The limiting belief {\em of each receiving agent} is always degenerate, meaning that it collapses to a single hypothesis, when sufficient time for learning is allowed.~\hfill$\square$
\end{remark}
\begin{remark}[Discord]
Different agents can in principle be in {\em discord}, since they can converge to different hypotheses. 
The particular behavior (who chooses what) will depend on a weighted combination of KL divergences.
~\hfill$\square$
\end{remark}
\begin{remark}[Mind Control]
We see from~(\ref{eq:avDiv}) and~(\ref{eq:uniquemin}) that only the local divergences corresponding to the sending agents, $\ell\in\mathcal{S}$, determine the value of $\mathscr{D}_k(\theta)$ and, hence, of $\theta^{\star}_k$.
Therefore, the limiting hypothesis $\theta^{\star}_k$ at agent $k$ is determined by the KL divergences pertaining only to the statistical models {\em within the sending sub-networks}, and, hence, {\em irrespective of the data sensed at agent $k$ within its receiving sub-network.} 
In a nutshell, we see the emergence of a mind-control effect: $i)$ the final states of the receiving agents are dependent only upon the properties of the detection problems at the {\em sending} agents; and $ii)$ different network topologies allow the sending agents to drive the receiving agents to potentially different decisions.
The emergence of a mind-control effect over weakly-connected networks was already discovered in~\cite{YingSayed2016,Salami}. 
Here, we establish a similar effect albeit one where the receiving agents attain degenerate beliefs. In comparison, in~\cite{YingSayed2016,Salami}, receiving agents end up assigning nonzero probabilities to more than one belief.~\hfill$\square$
\end{remark}

\begin{remark}[Distinctions relative to~\cite{Salami}]
Besides these commonalities, there are nevertheless important distinctions between the behavior observed in our setting and what was observed in~\cite{Salami}. 
First, for the linear-belief-combination algorithm used in~\cite{Salami}, the limiting belief of a receiving agent was shown to be a convex combination of the limiting beliefs of the sending agents, with the convex weights coming from the matrix $W$ in~(\ref{eq:omegadef}). This means that if two sending sub-networks have, e.g., limiting beliefs that collapse to different hypotheses, then the limiting belief of a receiving agent can have nonzero values a these two different locations. In comparison, in the log-belief-combination algorithm considered here the limiting beliefs are always concentrated at a {\em single} hypothesis.

Moreover, in~\cite{Salami}, the analysis required some regularity assumptions called {\em all-truths-are-equal} and {\em prevailing-signal} assumptions. These assumptions are not required in Theorem~\ref{theor:limbelief}. In a sense, the lack of these assumptions ascertains that some relevant effects, such as mind control, hold under greater generality and more relaxed settings.

Finally, it is useful to observe that one fundamental role in our setting is played by the weighting matrix $\Omega=E W$, while in~\cite{Salami} the main role was played by $W$ alone.~\hfill$\square$
\end{remark}

\subsection{Canonical Examples}
In order to examine in more detail the implications of Theorem~\ref{theor:limbelief}, we consider a simple yet insightful example. 
The sending and receiving components are:
\beq
\mathcal{S}=\mathcal{N}_1\cup\mathcal{N}_2,\qquad \mathcal{R}=\mathcal{N}_3,
\eeq
namely, we have two sending sub-networks, $\mathcal{N}_1$ and $\mathcal{N}_2$, and one receiving sub-network $\mathcal{N}_3$. 

For what concerns the inferential model, we assume there are three possible hypotheses, $\theta\in\{1,2,3\}$. 
The likelihood functions are the same across {\em all} agents.  
In particular, we assume that, for all $\xi\in\mathbb{R}$, and for $\theta\in\{1,2,3\}$:
\beq
L(\xi|\theta)=\frac{1}{\sqrt{2\pi}}\exp\left\{-\frac{(\xi-\mathsf{m}_\theta)^2}{2}\right\},
\eeq
where the means corresponding to the different hypotheses are chosen as, for some $\Delta>0$:% (see Fig.~\ref{fig:likeliexample})
\beq
\mathsf{m}_1=-\Delta,~~\mathsf{m}_2=0,~~\mathsf{m}_3=+\Delta.
\label{eq:expeclikeli}
\eeq
We further assume that the {\em true} distributions of the sending sub-networks (recall that only the sending sub-networks determine the limiting beliefs of the receiving sub-network) are Gaussian distributions, with expectations chosen among the expectations in~(\ref{eq:expeclikeli}). In particular, we assume that agents belonging to sub-network $\mathcal{N}_1$ generate data according to model $\theta=1$, i.e., with expectation equal to $-\Delta$, whereas agents belonging to sub-network $\mathcal{N}_2$ generate data according to model $\theta=3$, i.e., with expectation equal to $+\Delta$. Formally we write:
\beqa
f_\ell(\xi)&=&\frac{1}{\sqrt{2\pi}}\exp\left\{-\frac{(\xi+\Delta)^2}{2}\right\}, \quad \forall \ell\in\mathcal{N}_1,\\
f_\ell(\xi)&=&\frac{1}{\sqrt{2\pi}}\exp\left\{-\frac{(\xi-\Delta)^2}{2}\right\}, \quad \forall \ell\in\mathcal{N}_2.
\eeqa
Recalling that the KL divergence between two unit-variance Gaussian distributions of expectations $a$ and $b$ is given by $0.5 (a-b)^2$, under the setting described above we can write, for all $k\in\mathcal{R}$:
\beqa
\mathscr{D}_k(\theta)&=&\sum_{\ell\in\mathcal{S}}\omega_{\ell k}D[f_{\ell}||L_{\ell}(\theta)]\nonumber\\
&=&
\sum_{\ell\in\mathcal{N}_1} \omega_{\ell k} D[f_{\ell}||L(\theta)]
+
\sum_{\ell\in\mathcal{N}_2} \omega_{\ell k} D[f_{\ell}||L(\theta)]\nonumber\\
&=&
\frac{(-\Delta-\mathsf{m}_\theta)^2}{2}\sum_{\ell\in\mathcal{N}_1} \omega_{\ell k} 
+
\frac{(\Delta-\mathsf{m}_\theta)^2}{2}\sum_{\ell\in\mathcal{N}_2} \omega_{\ell k},\nonumber\\
\eeqa
which further implies:
\beqa
\mathscr{D}_k(1)&=&2\Delta^2\sum_{\ell\in\mathcal{N}_2} \omega_{\ell k}, 
\nonumber\\
\mathscr{D}_k(2)&=&\frac{\Delta^2}{2},
\nonumber\\
\mathscr{D}_k(3)&=&2\Delta^2\sum_{\ell\in\mathcal{N}_1} \omega_{\ell k},
\eeqa
where, in the intermediate equality, we used~(\ref{eq:convexomega}). 
As a result, we can compute the limiting hypothesis, for each $k\in\mathcal{R}$, as:
\beq
\theta^{\star}_k=\arg\!\min\left\{
4\sum_{\ell\in\mathcal{N}_2} \omega_{\ell k},~~1,~~ 4\sum_{\ell\in\mathcal{N}_1} \omega_{\ell k}
\right\}. 
\label{eq:argminexample}
\eeq
From~(\ref{eq:Omegaexpand}), one can argue that $\sum_{\ell\in \mathcal{N}_s}\omega_{\ell k}$ reflects the sum of influences over {\em all} paths connecting {\em all} sending agents in sub-network $s$ to receiving agent $k$. 

In order to find the minimizer in~(\ref{eq:argminexample}), we start by using~(\ref{eq:convexomega}) in~(\ref{eq:argminexample}), which yields:
\beq
\theta^{\star}_k=\arg\!\min\left\{
1-\sum_{\ell\in\mathcal{N}_1} \omega_{\ell k},~~0.25,~~ 
\sum_{\ell\in\mathcal{N}_1} \omega_{\ell k}
\right\}. 
\label{eq:argminexampleconvex}
\eeq
In view of Theorem~\ref{theor:limbelief}, the belief of the $k$-th receiving agent will converge to $\theta^{\star}_k=1$ if the following two conditions are simultaneously verified:
\beq
\begin{array}{ll}
1-\displaystyle{
\sum_{\ell\in\mathcal{N}_1} \omega_{\ell k}< 0.25
}
&\Leftrightarrow ~~
\displaystyle{\sum_{\ell\in\mathcal{N}_1} \omega_{\ell k}>0.75},
\\
\\
1-\displaystyle{\sum_{\ell\in\mathcal{N}_1} \omega_{\ell k}<\sum_{\ell\in\mathcal{N}_1} \omega_{\ell k}}
&\Leftrightarrow ~~\displaystyle{\sum_{\ell\in\mathcal{N}_1} \omega_{\ell k}>0.5.}
\end{array}
\label{eq:omegaineq}
\eeq
Taking the most stringent condition in~(\ref{eq:omegaineq}) reveals that:
\beq
\theta^{\star}_k=1 ~\Leftrightarrow~ \sum_{\ell\in \mathcal{N}_1}\omega_{\ell k}>0.75.
\label{eq:dominant1}
\eeq
In summary, we conclude that agent $k$ follows the opinion promoted by sending sub-network $\mathcal{N}_1$ if the influence of sub-network $\mathcal{N}_1$ on agent $k$ is ``sufficiently large''.

The situation is reversed if the influence of sub-network $\mathcal{N}_2$ is sufficiently large, namely, 
\beq
\theta^{\star}_k=3 ~\Leftrightarrow~ \sum_{\ell\in \mathcal{N}_2}\omega_{\ell k}>0.75,
\label{eq:dominant2}
\eeq
where we recall that hypothesis $\theta=3$ is promoted by sub-network $\mathcal{N}_2$.
However, there is another possibility. It occurs when:
\beq
\sum_{\ell\in \mathcal{N}_1}\omega_{\ell k}<0.75~~\text{ and }~~\sum_{\ell\in \mathcal{N}_2}\omega_{\ell k}<0.75.
\label{eq:nodominant}
\eeq
In this case, no clear dominance from one sub-network can be ascertained, and each receiving agent will choose $\theta^{\star}_k=2$, i.e., an {\em opinion that does not coincide with any of the opinions promoted by the sending sub-networks}. 

From~(\ref{eq:dominant1}) and~(\ref{eq:dominant2}), we see that the dominance of one of the sending sub-networks is determined by the aggregate influence $\sum_{\ell\in \mathcal{N}_1}\omega_{\ell k}$, with the complementary aggregate influence being $\sum_{\ell\in \mathcal{N}_2}\omega_{\ell k}=1-\sum_{\ell\in \mathcal{N}_1}\omega_{\ell k}$.
The main way to manipulate these factors consists in varying the {\em sizes} of the sending sub-networks or their connections with the receiving agents. 

In order to illustrate more carefully the possible scenarios, we consider the following simulation framework:
\begin{itemize}
	\item The strongly-connected sending components $\mathcal{N}_1$ and $\mathcal{N}_2$ are generated as Erd\H{o}s-R\'enyi random graphs with connection probability $q$, and the entries of the corresponding combination matrix are determined by the averaging rule, namely,\footnote{When drawing the random graph, we have verified that there exists at least one self-loop.}
	\begin{equation}
		a_{\ell k}=\begin{cases}
		1/n_k, &\text{ if $k\neq \ell$ are neighbors or $k=\ell$}\\
		0, &\text{ otherwise}
		\end{cases}
	\end{equation}
	where $n_k$ is the number of neighbors of node $k$ (including node $k$ itself). 
	In our experiments we set $q=0.7$.
	\item An agent $k$ is connected to a sending agent through a Bernoulli distribution with parameter $\pi_s$, which depends on the sending sub-network $s$. 
	Given the total number $d_k$, of directed edges from sending agents to agent $k$, we initially set $a_{\ell k}=1/d_k$. The combination matrix $A$ of the overall network $\mathcal{N}_1\cup\mathcal{N}_2\cup\mathcal{N}_3$ is normalized so that it is left-stochastic.
\end{itemize}
It is now possible to examine different scenarios by manipulating the size of the sending sub-networks as well as the send-receive connection probabilities $\pi_s$. 

\begin{figure}[!htb]
\centering
\[
\begin{array}{cccc}
{\centering{ ~~~~~~\textnormal{{\bf network graph}}}\par\medskip}
\\
\includegraphics[width=55mm]{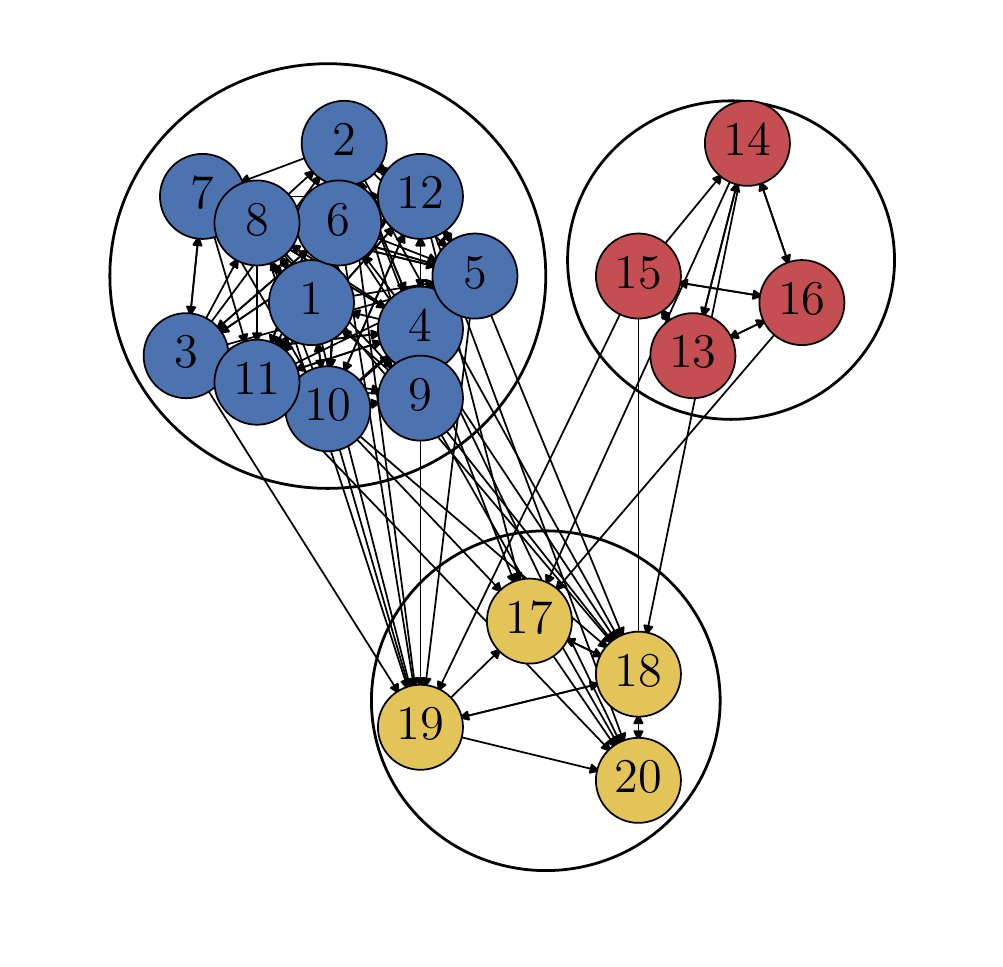}
\\
{\centering{ ~~~~~~\textnormal{{\bf belief evolution}}}\par\medskip}
\\
\includegraphics[width=65mm]{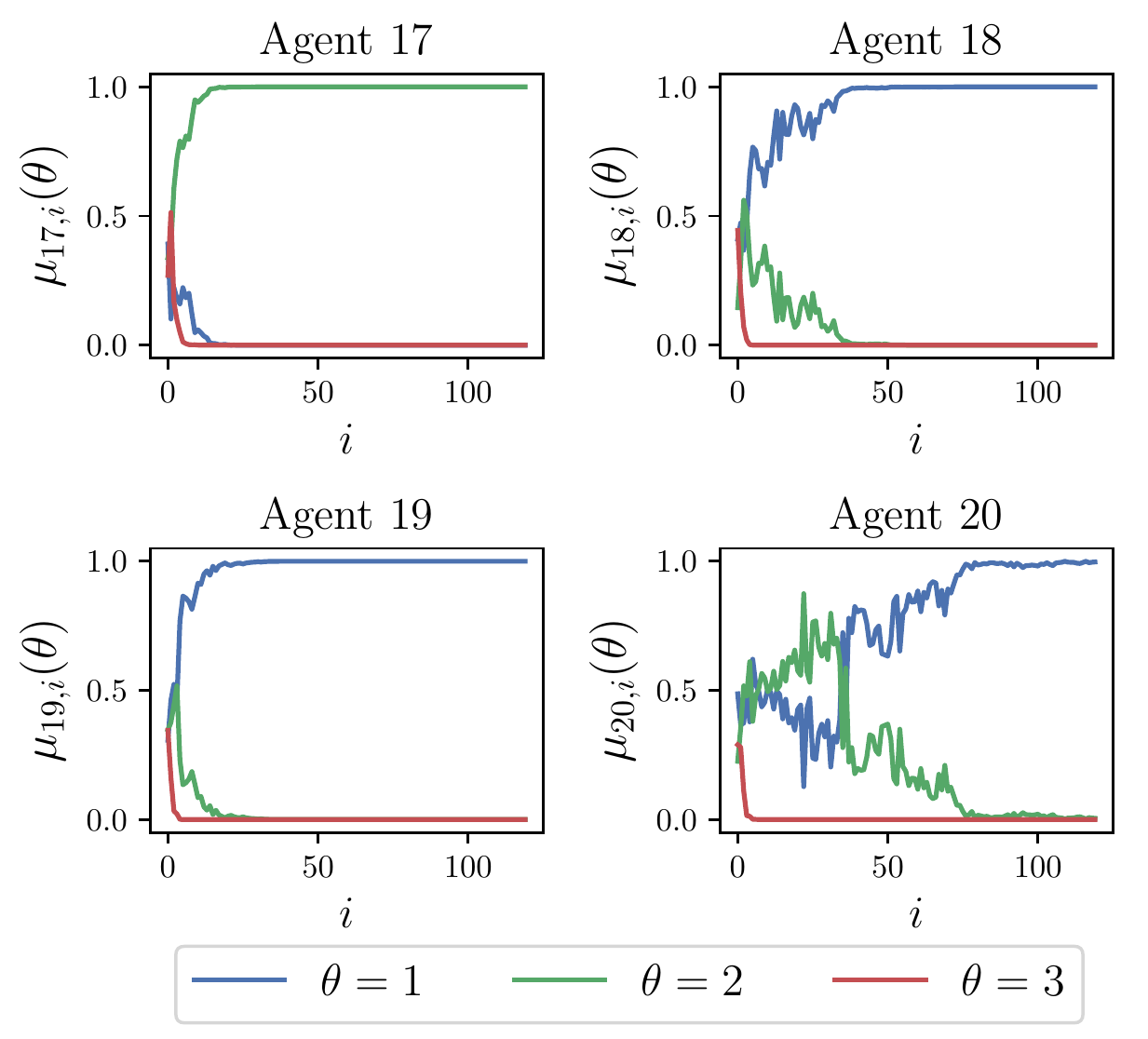}
\end{array}
\]
\caption{{\em How majorities build a majority}: Convergence of beliefs when the size of sending sub-network $\mathcal{N}_1$ is dominant.}
\label{fig:example1}
\end{figure}

\begin{figure}[!htb]
\centering
\[
\begin{array}{cccc}
{\centering{ ~~~~~~\textnormal{{\bf network graph}}}\par\medskip}
\\
\includegraphics[width=55mm]{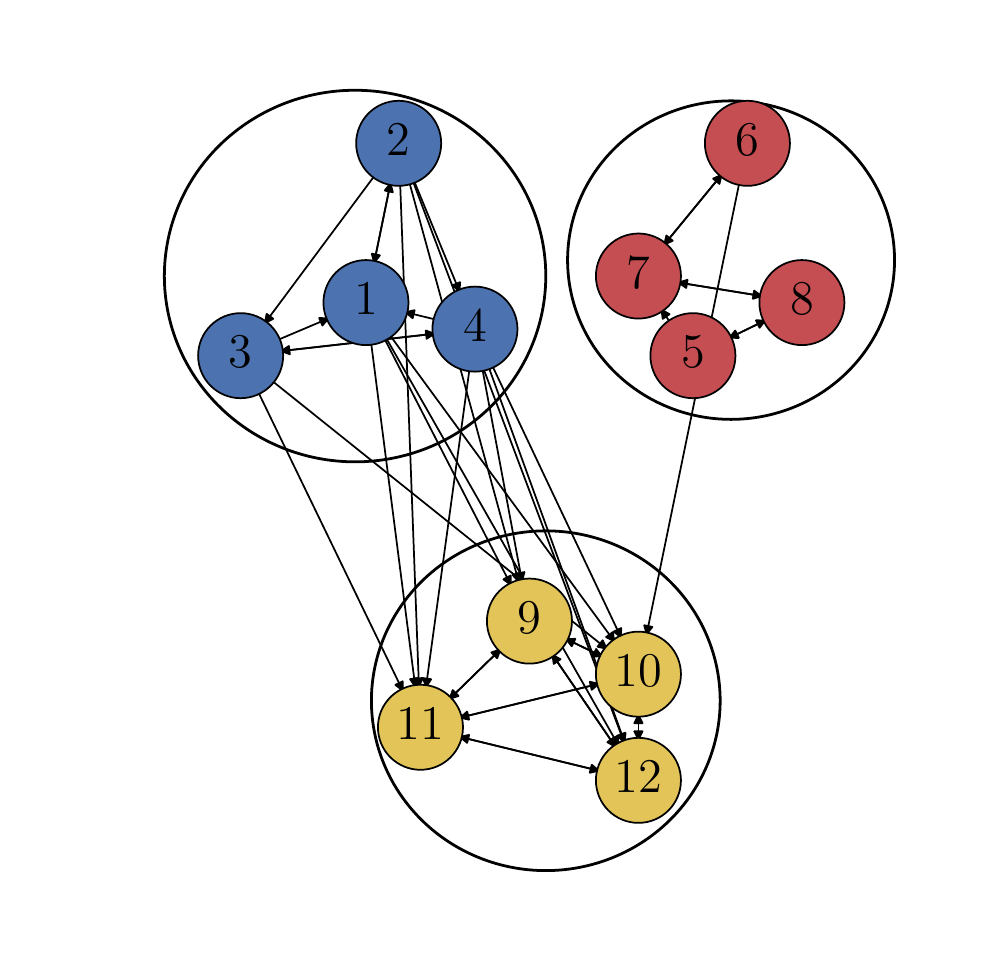}
\\
{\centering{ ~~~~~~\textnormal{{\bf belief evolution}}}\par\medskip}
\\
\includegraphics[width=65mm]{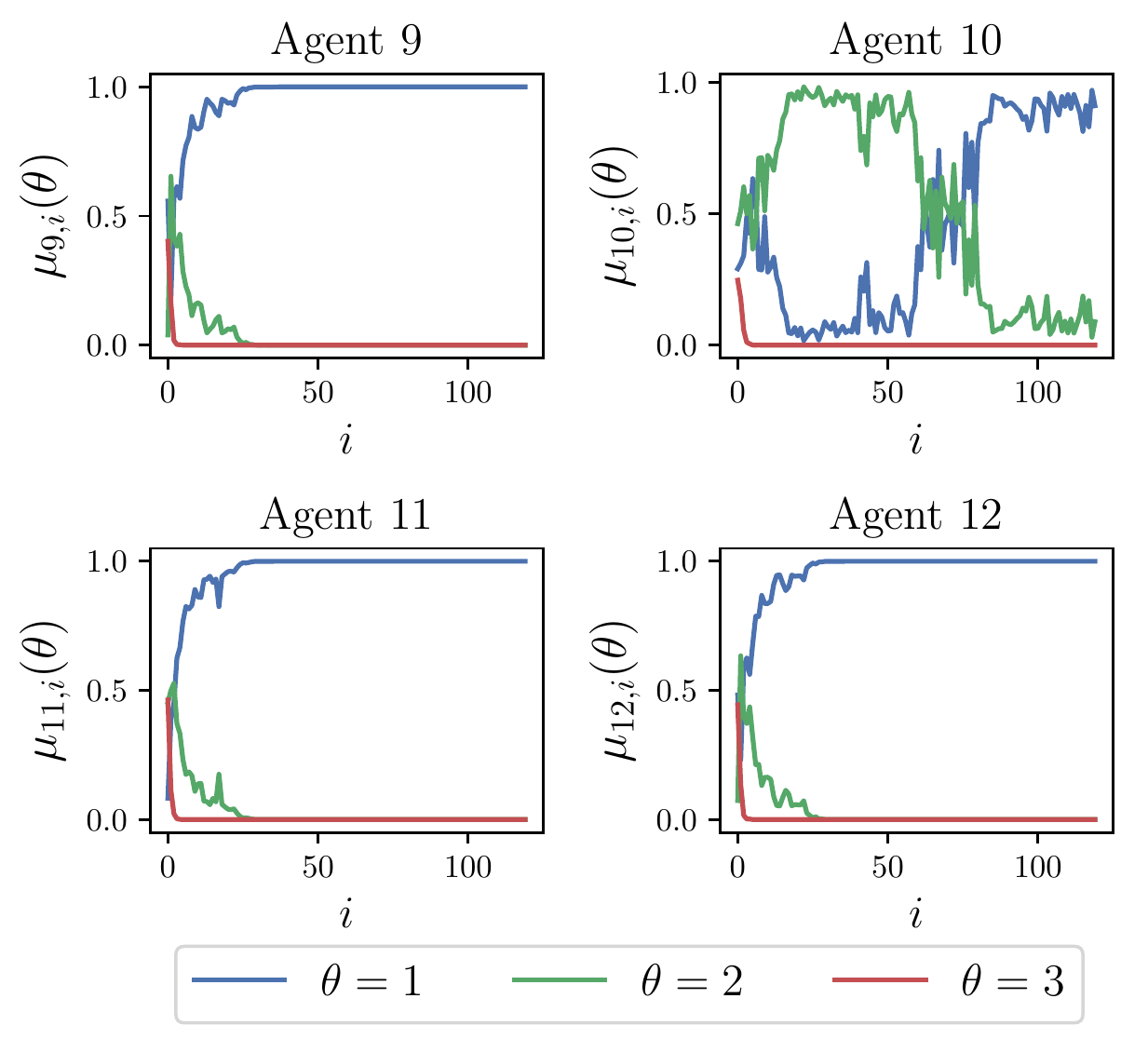}
\end{array}
\]
\caption{{\em How filter bubbles build a majority}: Convergence of beliefs when the connectivity from sending sub-network $\mathcal{N}_1$ is dominant.}
\label{fig:example2}
\end{figure}

\begin{figure}[!htb]
\centering
\[
\begin{array}{cccc}
{\centering{ ~~~~~~\textnormal{{\bf network graph}}}\par\medskip}
\\
\includegraphics[width=55mm]{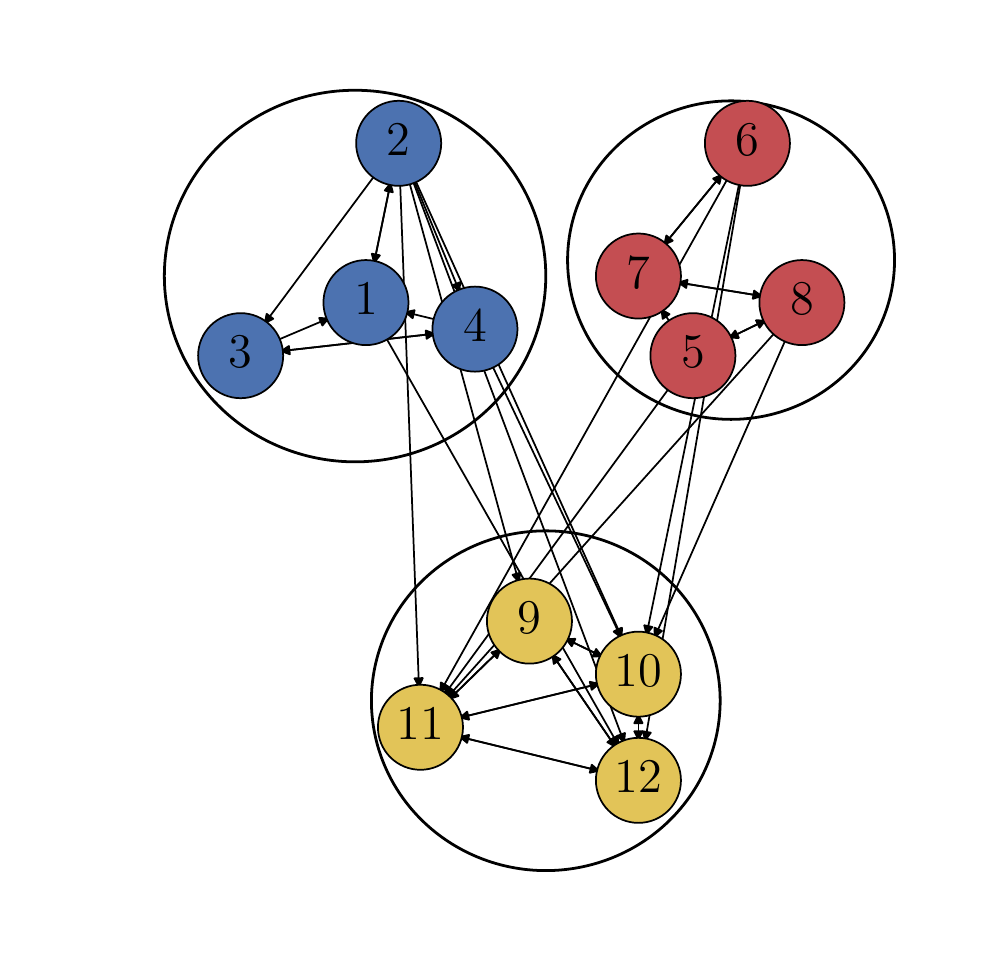}
\\
{\centering{ ~~~~~~\textnormal{{\bf belief evolution}}}\par\medskip}
\\
\includegraphics[width=65mm]{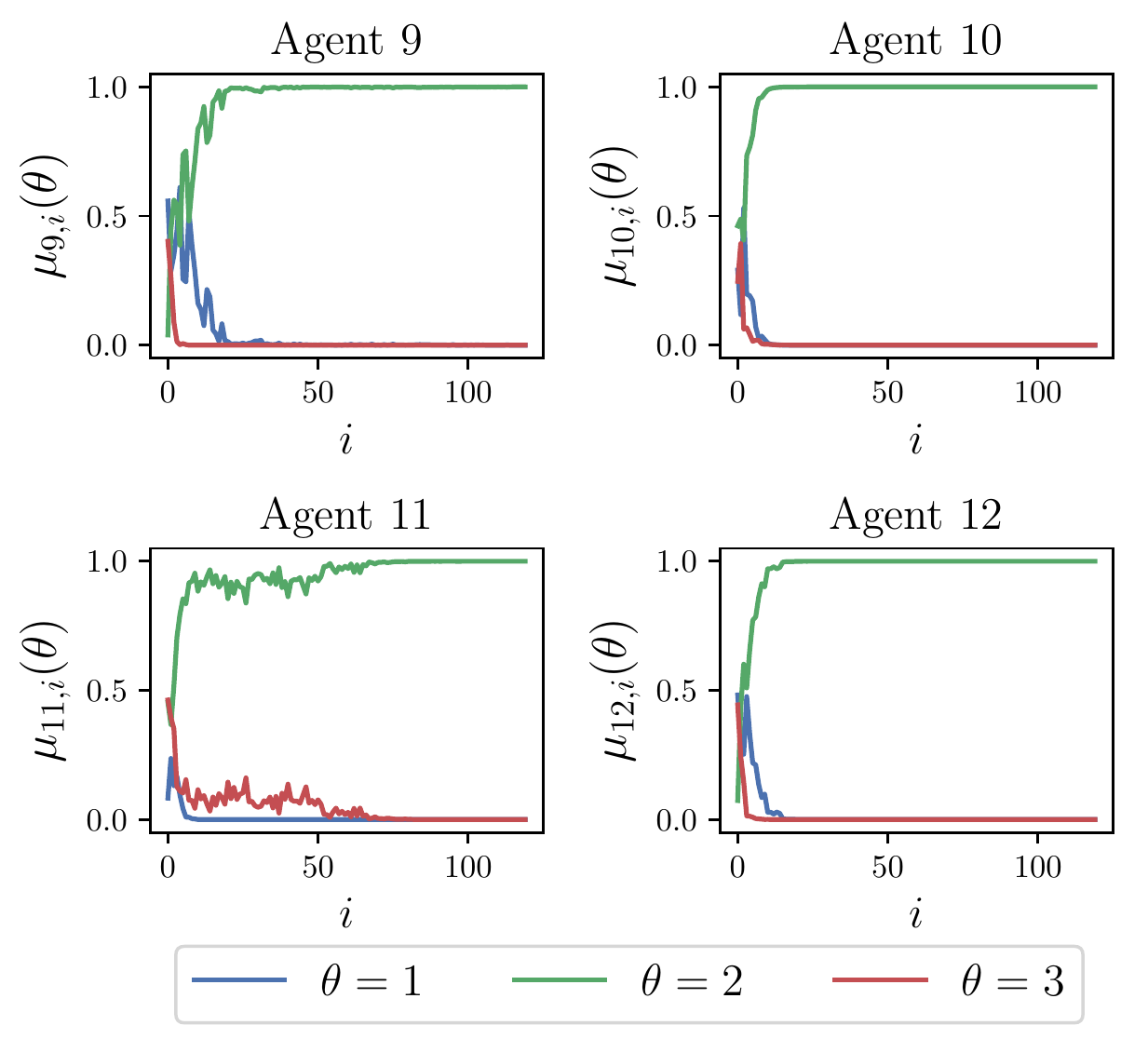}
\end{array}
\]
\caption{{\em Truth is somewhere in between}: Convergence of beliefs under balanced influences.}
\label{fig:example3}
\end{figure}

--- Setup $1$ or ``{\em How majorities build a majority}''. 
In Fig.~\ref{fig:example1}, we set $\pi_1=\pi_2=0.5$, i.e. it is equally probable that a receiving agent connects to any sending agent, irrespective of the sending sub-network. 
In view of this uniformity, we can expect that the limiting weights $\omega_{\ell k}$ are sufficiently uniform across the two sending sub-networks and, hence, that the value of $\sum_{\ell\in\mathcal{N}_1}\omega_{\ell k}$ is primarily determined by the sub-network size $N_1$.
In the example we are going to illustrate, we assume that the number of agents in sub-network $\mathcal{N}_1$ is three times larger than the size of sub-network $\mathcal{N}_2$. 
From the lowermost panel in Fig.~\ref{fig:example1}, we observe that receiving agents $(18,19,20)$ converge to $\theta=1$, i.e., to the opinion promoted by $\mathcal{N}_1$.
We see also that agent $17$ takes a minority position and opts for $\theta=2$, i.e., it does follow neither the opinion promoted by $\mathcal{N}_1$ nor by $\mathcal{N}_2$. 
This shows the following interesting effect. Even if sub-network $\mathcal{N}_1$ is bigger, for the specific topology shown in the example (see uppermost panel of Fig.~\ref{fig:example1}), the aggregate weight of agent $17$ is $\sum_{\ell\in\mathcal{N}_1}\omega_{\ell \, 17}=0.645$. This means that condition~(\ref{eq:nodominant}) is actually verified, which explains why agent $17$ opts for $\theta=2$.
In summary, we observed that building a majority of agents in $\mathcal{N}_1$ relative to $\mathcal{N}_2$ yields a majority of receiving agents opting for the hypothesis promoted by $\mathcal{N}_1$.

--- Setup $2$ or ``{\em How filter bubbles build a majority}''. 
Under this setup, we assume that both sending components have the same size, however $\pi_s$ is different for each of the two components. 
We set $\pi_1=0.9$ and $\pi_2=0.1$ in order to motivate agent $k$ to have more connections with sub-network $\mathcal{N}_1$ than with $\mathcal{N}_2$. This scenario is considered in Fig.~\ref{fig:example2}, where we see that all agents end up agreeing with opinion $\theta=1$, i.e., with the opinion promoted by the  sending component $\mathcal{N}_1$. 
Therefore, closing a receiving agent into the ``{\em filter bubble}'' determined by the overwhelming flow of data coming from $\mathcal{N}_1$ essentially makes these agents blind to the solicitations coming from $\mathcal{N}_2$. 
We notice that, while in the example of Fig.~\ref{fig:example1} one receiving agent behaves differently from the majority of agents, for the specific parameters used in Fig.~\ref{fig:example2} {\em all} agents opt for the same hypothesis. 
However, another type of distinction arises in terms of convergence rate. We observe that agent $10$ is reluctant to the received solicitations, since for the first half of the observation window the preferred hypothesis is $\theta_2$, and the convergence to $\theta_1$ is significantly slower than the convergence of the other agents.

--- Setup $3$ or ``{\em Truth is somewhere in between}''.
We now address the balanced case where the sending sub-networks have the same size and similar number of connections to the receiving sub-network ($\pi_1=\pi_2=0.5$). 
Under this setting, it is expected that no dominant behavior emerges, and~(\ref{eq:nodominant}) holds. 
We see in Fig.~\ref{fig:example3} that the receiving agents' opinions tend to converge with full confidence to hypothesis $\theta=2$ ($\mathsf{m}_\theta=0$), which is an opinion pushed by none of the sending agents. How can we explain this effect? 
One interpretation is that, in the presence of {\em conflicting suggestions} coming from the two sub-networks, the receiving agent opts for a conservative choice. If sending sub-network $\mathcal{N}_1$ says ``choose $-\Delta$'', while sending sub-network $\mathcal{N}_2$ says ``choose $+\Delta$'', then the receiving agent prefers to be agnostic and stays in the middle, i.e., it chooses $0$. 
Referring to real-life situations, we can think of one person betting on a soccer match between teams A and B. 
Assuming that discordant solicitations come from the environment, i.e., the person receives data suggesting to bet on the victory of team A, as well as data suggesting to bet on the victory of team B. If there is no sufficient evidence to let one suggestion prevail, then the most probable choice would be betting on a draw!
This ``{\em truth-is-somewhere-in-between}'' effect is a remarkable effect that is peculiar to the weakly-connected setting, and that has been not observed before, e.g., it was not present in~\cite{Salami}.

In summary, it is the cumulative influence of a sending group over a receiving agent that determines whether it will follow the group's opinion or not. 
This situation emulates the social phenomenon of herd behavior: agents choose to ignore their private signal in order to follow the most influencing group of agents. 
When none of the above dominance situations occurs, the receiving agent can opt for an opinion that is not promoted by any of the sending agents.  

%%%%%%%%%%%%%%%%%%%%%%%%%%%%%%%%%%%%

\section{Topology Learning}
In the previous section we examined the effect of the network topology on the social learning of the agents. In particular, we discovered how the topology and the states of the sending agents determine the opinion formation by the receiving agents.
The way the information is delivered across the network ultimately determines the minimizers in~(\ref{eq:uniquemin}), i.e., the value that each receiving agent's belief will converge to. 
We now examine the reverse problem. Assume we observe the belief evolution of part of the network. We would like to use this information to infer the underlying influences and topology. This is a useful question to consider because understanding the topology can help us understand why a particular agent adopts a certain opinion. 
The main question we consider now is this: given some measurements collected at the receiving agents, can we estimate their connections to the sending sub-networks? 

We shall answer this question under the following assumption of homogeneity of likelihoods and true distributions inside the individual sending sub-networks.

\begin{assumption}[Homogeneity within sending sub-networks]
\label{assum:homogeneity}
For $s=1,2,\ldots,S$, we assume that the distribution and the likelihood functions within the $s$-th sending sub-network are equal across all agents in that sub-network, namely, for all $\ell\in\mathcal{N}_s$:
\beq
\boxed{
f_{\ell}=f^{(s)},\qquad L_{\ell}(\theta)=L^{(s)}(\theta)
}
\eeq
~\hfill$\square$
\end{assumption}
One main consequence of Assumption~\ref{assum:homogeneity} is that~(\ref{eq:avDiv}) becomes:
\beqa
\mathscr{D}_k(\theta)&=&
\sum_{\ell\in\mathcal{S}}
\omega_{\ell k}
D[f_{\ell}||L_{\ell}(\theta)]
\nonumber\\
&=&\sum_{s=1}^S 
\left(D[f^{(s)} || L^{(s)}(\theta)] \sum_{\ell\in\mathcal{N}_s} \omega_{\ell k}\right),
\label{eq:homogeneity}
\eeqa
where ${\cal N}_s$ denotes the collection of agents in the $s$-th sending sub-network. 
Equation~(\ref{eq:homogeneity}) has the following relevant implication. 
Under Assumption~\ref{assum:homogeneity}, the network topology influences the average divergence $\mathscr{D}_k(\theta)$ only through an {\em aggregate} weight:
\beq
\boxed{
x_{sk}\dfz\sum_{\ell\in\mathcal{N}_s}\omega_{\ell k}=\sum_{\ell\in\mathcal{N}_s}w_{\ell k}
}
\label{eq:aggregateweightdef}
\eeq
The latter equality, using $w_{\ell k}$ instead of $\omega_{\ell k}$, comes straightforwardly from~(\ref{eq:perron}) and~(\ref{eq:omegadef}). 
This equality reveals that the aggregate weights depend solely on the matrix $W$, and not on the matrix $E$ of Perron eigenvectors. In other words, the inner structure of the pertinent sending sub-network $s$ does not influence the aggregate weight $x_{sk}$. 
We notice that, while a combination weight $a_{\ell k}$ accounts for a {\em local, small-scale} pairwise interaction between agent $\ell$ and agent $k$, the aggregate weight $x_{sk}$ accounts for {\em macroscopic} topology effects, for two reasons. 
First of all, $x_{sk}$ is determined by the {\em limiting} weights $\omega_{\ell k}$, which embody not only direct connection effects between $\ell$ and $k$, but also effects {\em mediated} by multi-hop paths connecting $\ell$ and $k$. 
Second, from~(\ref{eq:aggregateweightdef}) we see that $x_{sk}$ embodies the {\em global} effect coming from all agents belonging to the $s$-th sending component. 
In other words, $x_{sk}$ is a measure of the effect from all agents in sending sub-network $s$ on agent $k$. 
Since, in view of Theorem~\ref{theor:limbelief}, the average divergence determines the behavior of the limiting belief, we conclude from~(\ref{eq:homogeneity}) that the network topology ultimately influences the particular hypothesis chosen by a receiving agent only through these {\em global} weights $\{x_{sk}\}$. 

\begin{figure}[!t]
	\centering
	\includegraphics[width=.37\textwidth]{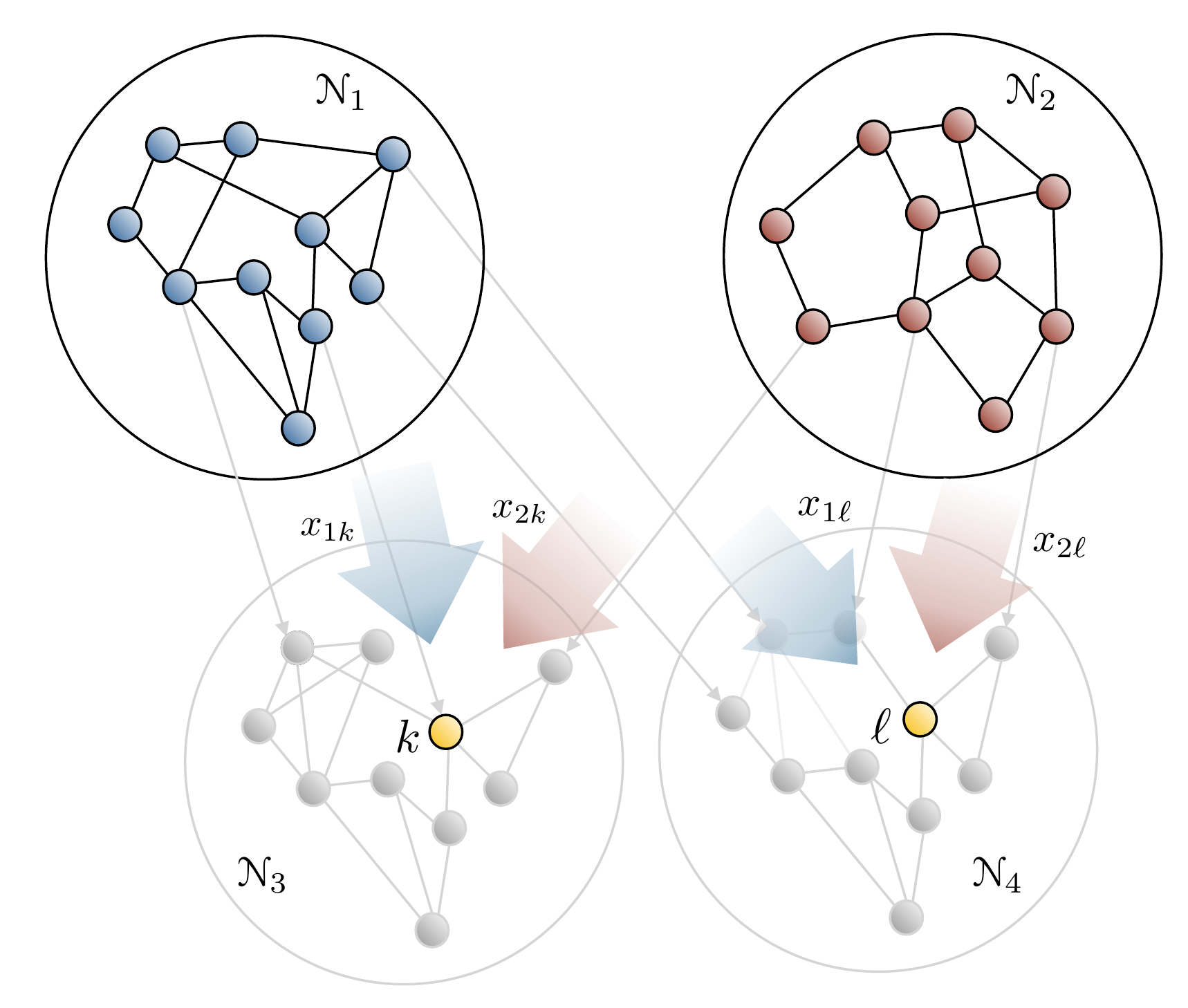}
	\caption{{\em Macroscopic} topology inference problem. The object of topology inference is constituted by the {\em global} weights $x_{s k}$ from sending sub-network $s$ to receiving agent $k$. For example, the weight $x_{1k}$ in the figure embodies the influence of {\em all} sending agents in $\mathcal{N}_1$, from {\em all} paths (possibly including intermediate receiving agents) leading to receiving agent $k\in\mathcal{N}_3$.}
	\label{fig:macrotopology}
\end{figure}

We assume that the data available for estimating $x_{sk}$ are the shared (intermediate) beliefs, $\bm{\psi}_{k,i}(\theta)$. 
We will say that {\em consistent} topology learning is achievable if the $x_{sk}$ can be correctly guessed when sufficient time is given for learning, i.e., we will focus on the {\em limiting} data, for all $\theta\neq\theta^{\star}_k$:\footnote{We remark that, in view of~(\ref{eq:psi}) in Appendix~\ref{app:proofTheor1}, the asymptotic properties of $\bm{\psi}_{k,i}(\cdot)$ are the same as $\bm{\mu}_{k,i}(\cdot)$.}
\beq
y_k(\theta)\dfz\lim_{i\rightarrow\infty} \frac{\log\bm{\psi}_{k,i}(\theta)}{i}\stackrel{\textnormal{a.s.}}{=} 
\mathscr{D}_{k}(\theta^{\star}_k) - \mathscr{D}_{k}(\theta).
\label{eq:zklimdef}
\eeq
Accordingly, the topology inference problem we are interested in can be formally stated as follows. 
For any receiving agent $k$, introduce its global-weight vector:
\beq
x_k\dfz[x_{1k},x_{2k},\ldots,x_{Sk}]^{\top},
\label{eq:xvec}
\eeq
and consider the vector stacking the $H$ limiting beliefs $y_k(\theta)$ (i.e., the data):
\beq
y_k\dfz [y_k(1),y_k(2),\ldots,y_k(H)]^{\top}.
\label{eq:yvec}
\eeq
The main question is whether we can estimate $x_k$ consistently from observation of $y_k$. 
In the sequel we will sometimes refer to this problem as a {\em macroscopic} topology inference problem --- see Fig.~\ref{fig:macrotopology} for an illustration.

As compared to other topology inference problems, we are faced here with one critical element of novelty. We have no data coming from the sending agents. This means that correlation between sending and receiving agent pairs cannot be performed. 
This is in sharp contrast with traditional topology inference problems, where the estimation of connections between pairs of agents is heavily based on comparison (e.g., correlation) between data streams coming from these {\em pairs} of agents~\cite{tomo,SantosMattaSayedIT2019,mateos}.  
In contrast, we focus here on the asymmetrical case that, when estimating the weights $x_{sk}$ from sending to receiving agents, no data are available from the sending agents. 
For this reason, the topology learning problem addressed in this work is significantly different from other traditional topology problems studied in the literature.

\section{Is Macroscopic Topology Learning Feasible?}
We now examine the feasibility of the topology learning problem illustrated in the previous section. 

Let us preliminarily introduce a matrix $D=[d_{\theta s}]$, which collects the $H\times S$ divergences between any true distribution in the sending sub-networks and any likelihood, and whose $(\theta,s)$-th entry is:
\beq
[D]_{\theta s}=d_{\theta s}=D[f^{(s)}||L^{(s)}(\theta)].
\label{eq:Dmatdef}
\eeq
Using~(\ref{eq:xvec}) and~(\ref{eq:Dmatdef}) in~(\ref{eq:homogeneity}), the network divergence of receiving agent $k$, evaluated at $\theta$, can be written as: 
\beq
\mathscr{D}_k(\theta)=
\sum_{s=1}^S 
d_{\theta s} x_{s k}.
\label{eq:networkdivhomogen}
\eeq
Through~(\ref{eq:yvec}) we can rewrite the limiting data in~(\ref{eq:zklimdef}) as:
\beq
y_k(\theta)=\mathscr{D}(\theta^{\star}_k)-\mathscr{D}(\theta)=\sum_{s=1}^S (d_{\theta^{\star}_k s} - d_{\theta s}) \, x_{sk}.
\eeq
It is useful to introduce the matrix:
\beq
B_k\dfz \left(\mathbbm{1}_{H} e\T_{\theta^{\star}_k} - I_H\right) D,
\label{eq:Bkevec}
\eeq
where $e_m$ is an $H\times 1$ vector with all zeros and a one in the $m$-th position. 
It is important to note that $B_k$ has its $\theta^{\star}_k$-th row equal to zero. 
We can now formulate the topology problem in terms of the following constrained system:
\beq
\textnormal{Find }\widetilde{x}_k\in\mathbb{R}^S ~~~\textnormal{such that }
\left\{
\begin{array}{ll}
y_k=B_k \, \widetilde{x}_k, 
\vspace*{5pt}
\\
\sum_{s=1}^S \widetilde{x}_{sk}=1, 
\vspace*{5pt}
\\
\widetilde{x}_{k}>0,
\end{array}
\right.
\label{eq:linsystopology0}
\eeq
where we remark that the notation $\widetilde{x}_k>0$ signifies that all entries in the solution vector $\widetilde{x}_k$ must be strictly positive. 
This positivity constraint is enforced because by assumption, each receiving sub-network is connected to at least one agent from each sending sub-network, which implies that the true vector we are looking for, $x_k$, has all positive entries. 
The equality constraint in~(\ref{eq:linsystopology0}) can be readily included in matrix form by introducing the augmented matrix and vector:
\beq
C_k
\dfz 
\begin{bmatrix}
B_k
\\
\mathbbm{1}_S\T
\end{bmatrix},\qquad
\widetilde{y}_k\dfz
\begin{bmatrix}
y_k
\\
1
\end{bmatrix},
\label{eq:augmentedB}
\eeq
which allow rewriting~(\ref{eq:linsystopology0}) as:
\beq
\boxed{
\textnormal{Find }\widetilde{x}_k\in\mathbb{R}^S: ~~\widetilde{y}_k=C_k\, \widetilde{x}_k,~~~\widetilde{x}_k>0
}
\label{eq:linsystopology}
\eeq 
We are now ready to state formally the concept of feasibility for the topology learning problem. 
First, we want to solve the problem under the assumption that the matrix of divergences, $D$, is known, i.e., that sufficient knowledge is available about the underlying statistical models (likelihoods and true distributions).
In this respect, we remark that the matrix $B_k$ in~(\ref{eq:Bkevec}) depends on $\theta^{\star}_k$, which in turn depends on the unknowns $x_{sk}$ as well through~(\ref{eq:uniquemin}).  
However, from Theorem~\ref{theor:limbelief} we know that the beliefs (and also the intermediate beliefs) converge to $1$ at $\theta^{\star}_k$. Therefore, we can safely estimate $\theta^{\star}_k$ from the limiting data $y_k(\theta)$, which is tantamount to assuming that the matrix $B_k$ is known. 

Therefore, achievability of a consistent solution for the topology learning problem translates into the condition that the linear system in~(\ref{eq:linsystopology}) should admit a unique solution. We will now prove the following result.
 
\begin{lemma}[Necessary Condition for Macroscopic Topology Learning]
\label{prop:necessary}
The topology learning problem described by the system in~(\ref{eq:linsystopology}) admits a unique solution if, and only if:
\beq
\boxed{
\mathrm{rank}(C_k)=S
}
\label{eq:necsufcond}
\eeq
Thus, a necessary condition for topology learning is that the number of hypotheses is at least equal to the number of sending sub-networks, namely, that: 
\beq
\boxed{
H\geq S
}
\label{eq:necond}
\eeq~\hfill$\blacksquare$
\end{lemma}
\begin{IEEEproof}
We remark that we are not concerned with the existence of a solution for the constrained linear system~(\ref{eq:linsystopology}). In fact, this system admits at least a solution, namely, the true weight vector, $x_k\in\mathbb{R}_{+}^S$, which by assumption fulfills the equation $\widetilde{y}_k=C_k \, x_k$.

Let us now focus on the unconstrained system (i.e., the system in~(\ref{eq:linsystopology}) {\em without the inequality constraints}), whose set of solutions is given by~\cite{horn2012matrix}:
\beq
\widetilde{x}_k=C_k^{\dagger}\,\widetilde{y}_k + (I_S - C_k^{\dagger}C_k)z, 
\label{eq:genlinsystform}
\eeq
where $z\in\mathbb{R}^S$ is an arbitrary vector, and $C_k^{\dagger}$ is the Moore-Penrose pseudoinverse of $C_k$. 
If $\mathrm{rank}(C_k)=S$, it is well known~\cite{horn2012matrix} that $C_k^{\dagger}=(C_k\T C_k)^{-1} C_k\T$, which implies that the second term on the RHS in~(\ref{eq:genlinsystform}) is zero, which in turn implies that the unconstrained system has the unique solution:
\beq
\boxed{
\widetilde{x}_k=(C_k\T C_k)^{-1} C_k\T \widetilde{y}_k=x_k
}
\label{eq:systinvert}
\eeq
The latter equality holds because, if the unconstrained system has a {\em unique} solution, this is also the unique solution for the constrained system, i.e., it coincides with $x_k$ and satisfies the positivity constraints.
Accordingly, we have proved that whenever $\mathrm{rank}(C_k)=S$, the {\em constrained} system has the unique solution corresponding to the true vector $x_k$. 

We now show that when $\mathrm{rank}(C_k)<S$ the constrained system has infinite solutions. 
Since any solution of the unconstrained system takes on the form~(\ref{eq:genlinsystform}), and since $x_k$ {\em is} a particular solution, there will exist a certain vector $z_0$ such that the $x_k$ can be written as: 
\beq
x_k=C_k^{\dagger}\,\widetilde{y}_k + (I_S - C_k^{\dagger}C_k)z_0.
\eeq
Consider a solution $\widetilde{x}_k$ in~(\ref{eq:genlinsystform}) that corresponds to another vector, $z=z_0+\epsilon$, where $\epsilon$ is a perturbation vector:
\beq
\widetilde{x}_k=C_k^{\dagger}\,\widetilde{y}_k + (I_S - C_k^{\dagger}C_k)(z_0+\epsilon)=
x_k + (I_S - C_k^{\dagger}C_k)\epsilon.
\label{eq:perturbepsilon}
\eeq
Since by assumption $x_k>0$, we conclude from~(\ref{eq:perturbepsilon}) that for sufficiently small perturbations it is always possible to obtain a distinct $\widetilde{x}_k>0$, which implies that the {\em constrained} system in~(\ref{eq:linsystopology}) has infinite solutions.

In summary, we conclude that the topology learning problem is feasible if, and only if, $\mathrm{rank}(C_k)=S$. 
Finally, by observing that the augmented matrix $C_k$ is an $(H+1)\times S$ matrix with an all-zeros row, we have in fact proved the claim of the lemma.
\end{IEEEproof}

Lemma~\ref{prop:necessary} has at least three useful implications. 
%First, it reveals a fundamental interplay between social learning and topology learning, since it ascertains that the possibility of estimating the global weight vector $x_k$ is related to the interplay between an attribute of the social inferential problem (number of hypotheses $H$) and a topology attribute (number of sending sub-networks $S$).
First, it reveals a fundamental interplay between social learning and topology learning: the possibility of estimating $x_k$ depends on the comparison between two seemingly unrelated quantities, the number of hypotheses $H$ (an attribute of the social inferential problem) and the number of sending sub-networks $S$ (an attribute of the network topology).

Second, the necessary condition in~(\ref{eq:necond}) highlights that topology learning over social networks is challenging.
%, as its feasibility is not easily granted. 
For example, if the agents of the social network want to solve a binary detection problem ($H=2$), then the maximum number of sending sub-networks that could allow faithful topology estimation is $S=2$.
Increasing the complexity of the social learning problem %that the agents want to solve 
(i.e., increasing $H$) is beneficial to topology estimation, since it allows to increase also $S$.
%the maximum number of sending sub-networks.

Third, we see that having more sending sub-networks makes topology learning more complicated. This is because increasing the number of sending sub-networks increases the number of unknowns (i.e., the dimension of $x_k$), while not adding information since %we recall that 
in our setting we are not allowed to probe the sending nodes. Remarkably, when examining jointly the social learning and the topology learning problems, {\em the role of the data and of the unknowns is exchanged.} 
In the social learning problem, more hypotheses means more unknowns and more sending sub-networks means more data; in the topology learning problem, the situation is exactly reversed.

%%%%% I'M DELETING THIS COMMENT TO SAVE SPACE
%%%%%%%%%%%%%%%%%%%%%%%%%%%%%%
%%%%%%%%%%%%%%%%%%%%%%%%%%%%%%
%\textcolor{blue}{
%The condition $\mathrm{rank}(C_k)=S$ may appear to be overly restrictive. 
%One can attempt to relax the condition and solve~(\ref{eq:linsystopology}) under additional constraints. Typical constraints could be in terms of sparsity of the weights $x_{sk}$, and/or structural dependency among different receiving agents. 
%However, we prefer to address first the problem of achievability under the most stringent setting where such prior information is not available. 
%Moreover, we notice that in general: $i)$ the weights $x_{sk}$ might exhibit a low degree of sparsity since they are {\em global} weights; and $ii)$ the network topology does not necessarily impose strong correlations between the values $x_{sk}$ that could restrict the space of solutions for the system in~(\ref{eq:linsystopology}).}
%%%%%%%%%%%%%%%%%%%%%%%%%%%%%%
%%%%%%%%%%%%%%%%%%%%%%%%%%%%%%

\subsection{Structured Gaussian Models}
\label{sec:TLGauss}
In this section we consider the practical case of a Gaussian model, defined as follows. 
\begin{itemize}
\item
All agents use the same family of likelihood functions $\{L(\theta)\}$, for $\theta=1,2,\ldots,H$. 
\item
These likelihoods are unit-variance Gaussian likelihoods with different means $\{\mathsf{m}_\theta\}$. 
\item
Each true distribution coincides with one of the likelihoods. 
This implies that the distribution of the $s$-th sending sub-network, $f^{(s)}$, is a unit-variance Gaussian distribution with mean $\nu_s$ that is chosen among the means $\{\mathsf{m}_\theta\}$, namely, for $s=1,2,\ldots,S$: 
\beq
\nu_s\in\{\mathsf{m}_1,\mathsf{m}_2,\ldots,\mathsf{m}_H\}.
\eeq
\item
The sending sub-networks have different means. 
\end{itemize}
Using~(\ref{eq:Dmatdef}) and the definition of KL divergence between Gaussian distributions, the matrix $D$ is given by:
\beq
D=\frac 1 2
\begin{bmatrix}
	(\mathsf{m}_1-\nu_1)^2&(\mathsf{m}_1-\nu_2)^2&\dots &(\mathsf{m}_1-\nu_S)^2\\
	(\mathsf{m}_2-\nu_1)^2&(\mathsf{m}_2-\nu_2)^2&\dots &(\mathsf{m}_2-\nu_S)^2\\
	\vdots&&&\vdots\\
	(\mathsf{m}_H-\nu_1)^2&(\mathsf{m}_H-\nu_2)^2&\dots &(\mathsf{m}_H-\nu_S)^2
	\end{bmatrix}.
	\label{eq:EDMGaussGeneral}
\eeq
From~(\ref{eq:EDMGaussGeneral}) it is readily seen that, if the sending sub-networks share the same true distribution (i.e., if $\nu_1=\nu_2=\cdots=\nu_S$), then the matrix $D$ has rank $1$, and, hence, the topology learning problem is obviously not feasible. As said, we will instead focus on the opposite case where the true expectations are all distinct.

For ease of presentation, and without loss of generality we can assume that the sending sub-networks are numbered so that the expectations of the true distributions are:
\beq
\nu_1=\mathsf{m}_1, \,\nu_2=\mathsf{m}_2,\ldots,\nu_S=\mathsf{m}_S,
\eeq 
which implies that~(\ref{eq:EDMGaussGeneral}) takes on the form:
\beq
D=\frac 1 2
\begin{bmatrix}
	0&(\mathsf{m}_1-\mathsf{m}_2)^2&\dots &(\mathsf{m}_1-\mathsf{m}_S)^2\\
	(\mathsf{m}_2-\mathsf{m}_1)^2&0&\dots &(\mathsf{m}_2-\mathsf{m}_S)^2\\
	\vdots&&&\vdots\\
	(\mathsf{m}_H-\mathsf{m}_1)^2&(\mathsf{m}_H-\mathsf{m}_2)^2&\dots &(\mathsf{m}_H-\mathsf{m}_S)^2
	\end{bmatrix}.
	\label{eq:EDMGauss}
\eeq
The structure in~(\ref{eq:EDMGauss}) implies that, for $H=S$, the matrix $D$ is a Euclidean distance matrix (but for the constant $1/2$)~\cite{dokmanic2015euclidean}. 
These matrices are constructed as follows. 
Given points $r_1,r_2,\ldots,r_L$, belonging to $\mathbb{R}^{\sf dim}$, the $(i,j)$-th entry of the matrix ${\sf EDM}(r_1,r_2,\ldots,r_L)$ is given by the squared Euclidean distance between points $r_i$ and $r_j$. 
Accordingly, we see from~(\ref{eq:EDMGauss}) that, for $H=S$:
\beq
D=\frac 1 2 {\sf EDM}(\mathsf{m}_1,\mathsf{m}_2,\ldots,\mathsf{m}_H).
\eeq
For $H>S$, the matrix $D$ can be described as an {\em extended} Euclidean distance matrix, constructed as follows. 
Let:
\beqa
E_S&\dfz& \frac 1 2 {\sf EDM}(\mathsf{m}_1,\mathsf{m}_2,\ldots,\mathsf{m}_S),\nonumber\\
E_H&\dfz& \frac 1 2 {\sf EDM}(\mathsf{m}_1,\mathsf{m}_2,\ldots,\mathsf{m}_H),\nonumber\\
E_{H-S}&\dfz& \frac 1 2 {\sf EDM}(\mathsf{m}_{S+1},\mathsf{m}_{S+2},\ldots,\mathsf{m}_H),
\label{eq:variousEDMdef}
\eeqa
and let $F$ be the $(H-S)\times S$ matrix with entries, for $\theta=S+1,S+2,\ldots,H$ and $s=1,2,\ldots,S$:
\beq
[F]_{\theta s}=\frac 1 2 (\mathsf{m}_\theta-\mathsf{m}_s)^2.
\eeq
Then, we have the following representation:
\beq
D=\begin{bmatrix}
E_S\\
F
\end{bmatrix},\qquad
E_{H}=\begin{bmatrix}
E_S&F\T\\
F&E_{H-S}
\end{bmatrix}.
\label{eq:generalD}
\eeq
The following theorem, which establishes the feasibility of the topology learning problem for the considered Gaussian model, relies heavily on some fundamental properties of Euclidean distance matrices. 

\begin{theorem}[Macroscopic Topology Learning under Structured Gaussian Models]
\label{theor:TopologyGaussian}
Let $S\geq 2$ and $H\geq S$. 
Assume that all sending sub-networks have the same family of unit-variance Gaussian likelihood functions $L(\theta)$ with distinct means $\{\mathsf{m}_{\theta}\}$, for $\theta=1,2,\ldots,H$. 
Assume that the true distributions $f^{(s)}$, within the sending sub-networks $s=1,2,\ldots,S$, are unit-variance Gaussian with distinct means $\nu_s$, chosen from the collection $\{\mathsf{m}_{\theta}\}$. 
Then, under Assumption~\ref{assum:uniquemin} (so that the matrix $B_k$ in~(\ref{eq:Bkevec}) is well defined), for all receiving agents $k\in\mathcal{R}$ we have that:
\beq
\boxed{
\mathrm{rank}(C_k)=2
}
\label{eq:rank2}
\eeq
\end{theorem}
\begin{IEEEproof}
The proof is reported in Appendix~\ref{app:ProofTheor2}.
\end{IEEEproof}

\begin{remark}[Topology Learning under Structured Gaussian Models is Challenging]
In view of Lemma~\ref{prop:necessary}, Eq.~(\ref{eq:rank2}) has the following implication. 
Under the considered Gaussian model, topology learning is feasible only when $S=2$. We remark also that, when $S=2$, condition~(\ref{eq:necond}) plays no role, since any meaningful classification problem has at least $H=2$. 
In summary, Theorem~\ref{theor:TopologyGaussian} reveals that the structure of the Gaussian model makes topology learning very challenging, as this problem is not solvable for networks with more than $2$ sending sub-networks. 
Thus, the theorem reveals that $H\geq S$ is {\em not} a sufficient condition for consistent topology 
learning.~\hfill$\square$ 
\end{remark}

\subsection{Diversity Models}
\label{sec:TLrandomized}

We can now examine the effect that diversity in the models of the sending sub-networks can have on topology learning.
Since the limiting beliefs are essentially determined by the divergence matrix $D$, it is meaningful to impose a form of diversity in terms of the divergences between distributions and likelihoods. 
In other words, differently from the Gaussian case illustrated in the previous section, we now require that the entries of $D$ are not tightly related to each other, namely, we allow them to assume values in $\mathbb{R}_{+}^{H\times S}$ (where we denote by $\mathbbm{R}_+$ the nonnegative reals) with no strong structure linking them.   

One typical model for this type of diversity is that the divergences perceived by the different agents (i.e., across index $s$), and corresponding to different hypotheses (i.e., across index $h$), are modeled as absolutely continuous random variables. 
This randomness is a formal way to embody some degree of variability in how the agents ``see'' the world.  
For example, this is a useful model to consider when the agents, due to imperfect knowledge, have likelihoods that are slightly {\em perturbed} versions of some nominal model. Examples of this type are illustrated in the next section.

In order to avoid confusion, it is important to remark one fundamental property. 
Under the diversity setting, the matrix $\bm{D}$ is random\footnote{Accordingly, we will now use the bold notation for the matrix entries, $\bm{d}_{\theta s}$, as well as for other related quantities.} with entries modeled as absolutely continuous random variables. 
%and, hence, the hypothesis $\bm{\theta}^{\star}_k$ that minimizes the divergence is random as well. This implies that not only the nonzero entries, but also the location of the all-zeros row in $\bm{B}_k$ in~(\ref{eq:Bkevec}) is random.
The full-rank property for this type of matrices is a classical result. 
However, we observe from~(\ref{eq:Bkevec}) that the matrix $\bm{B}_k$ is obtained from $\bm{D}$ by multiplying a matrix that depends on a random variable $\bm{\theta}^{\star}_k$, which in turn depends statistically upon the entries of $\bm{D}$. 
Finally, we know from~(\ref{eq:augmentedB}) that $\bm{C}_k$ is obtained from $\bm{B}_k$ by adding an all-ones row. 
Accordingly, to determine the rank of $\bm{C}_k$ we need to address carefully these intricate dependencies.
This is accomplished in the proof of the forthcoming Theorem~\ref{theor:TopologyRandom}.

\begin{theorem}[Macroscopic Topology Learning under General Models with Diversity]
\label{theor:TopologyRandom}
Let $H\geq S$, and assume that the array $\{\bm{d}_{\theta s}\}$, with $\theta=1,2,\ldots,H$ and $s=1,2,\ldots,S$, is made of random variables that are jointly absolutely continuous with respect to the Lebesgue measure on $\mathbbm{R}_+^{H\times S}$. 
Then, for all receiving agents $k\in\mathcal{R}$ we have that, with probability $1$, Assumption~\ref{assum:uniquemin} is verified and the matrix $\bm{C}_k$ is full column rank, namely, 
\beq
\boxed{
\P\left[\bm{\theta}_k^{\star} \textnormal{ is unique and } \mathrm{rank}(\bm{C}_k)=S\right]=1
}
\eeq
\end{theorem}
\begin{IEEEproof}
The proof is reported in Appendix~\ref{app:ProofTheor3}.
\end{IEEEproof}
%%%%%%%%%

The meaning of Theorem~\ref{theor:TopologyRandom} is that {\em configurations of KL divergence that lead to a rank-deficient matrix $C_k$ are rare}.
In other words, if some diversity exists in the statistical models of the sending components, then the topology inference problem is feasible for almost all configurations.

\section{Simulation results}
We now present some illustrative examples. The first example refers to the Gaussian model presented in Sec.~\ref{sec:TLGauss}. The other two examples refer to the setting with diversity presented in Sec.~\ref{sec:TLrandomized}.

\def\paneldim{0.4}
\begin{figure}[t]
\begin{minipage}{.33\linewidth}
\vspace*{-20pt}
{\centering{\bf ~~~~~~network graph}\par\medskip}
%\hspace*{-20pt}
\centering
%\subfloat[]
{
\includegraphics[scale=\paneldim]{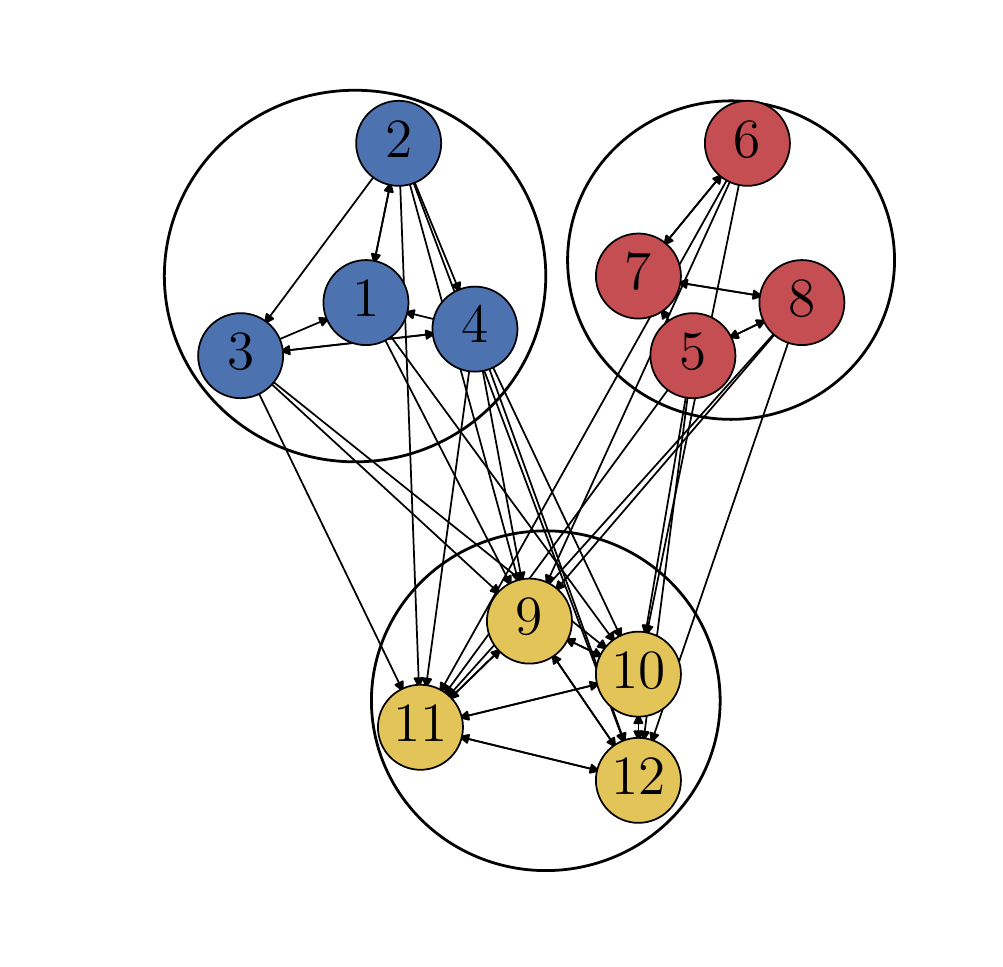}}
\end{minipage}
\begin{minipage}{.33\linewidth}
{\centering{\bf ~~~~~~belief evolution}\par\medskip}
%\vspace*{5pt}
%\hspace*{-10pt}
\centering
%\subfloat[]
{
\includegraphics[scale=\paneldim]{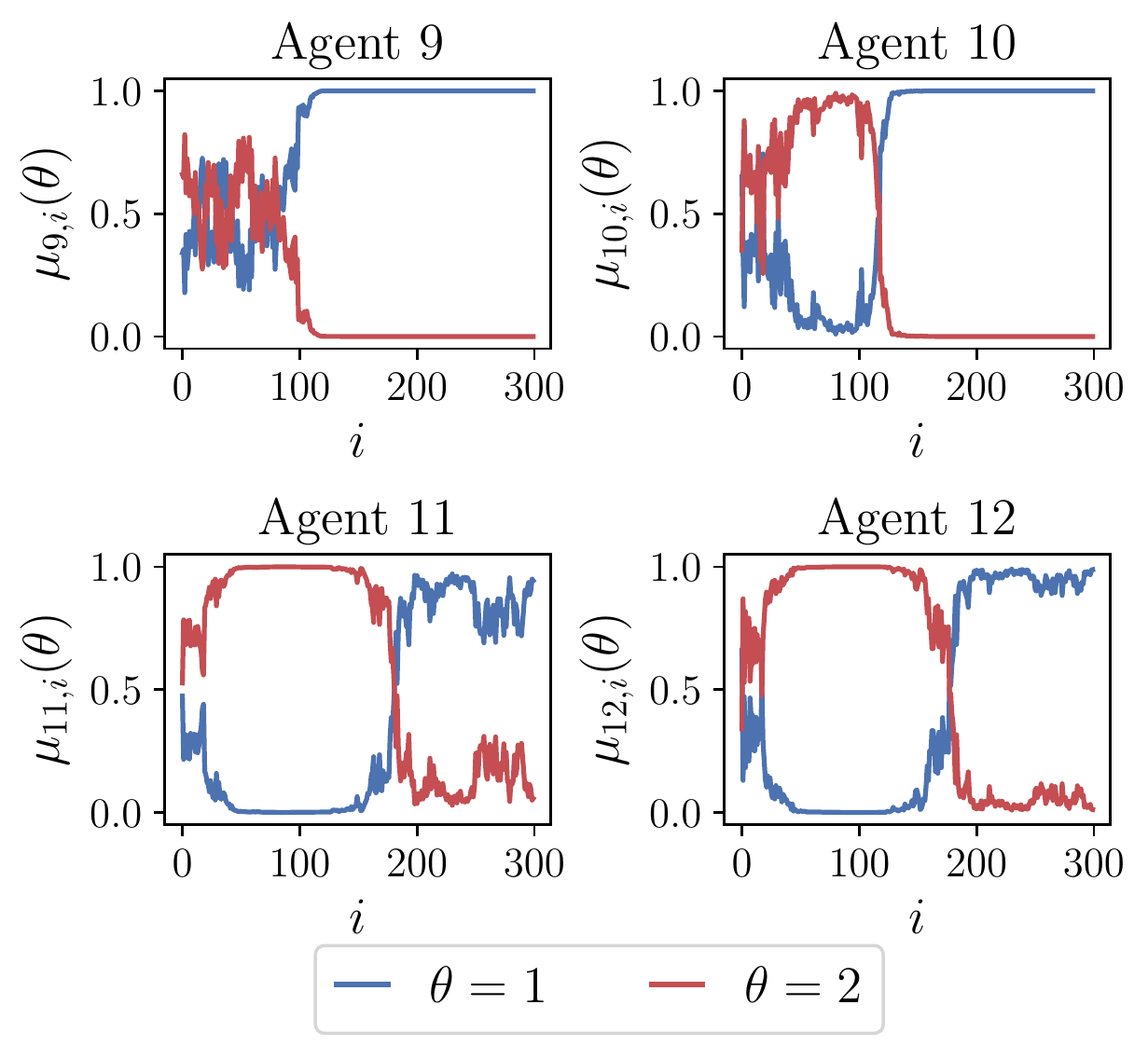}}
\end{minipage}
\begin{minipage}{.33\linewidth}
{\centering{\bf ~~~~~~estimated weights}\par\medskip}
%\vspace*{5pt}
\centering
%\subfloat[]
{
\includegraphics[scale=\paneldim]{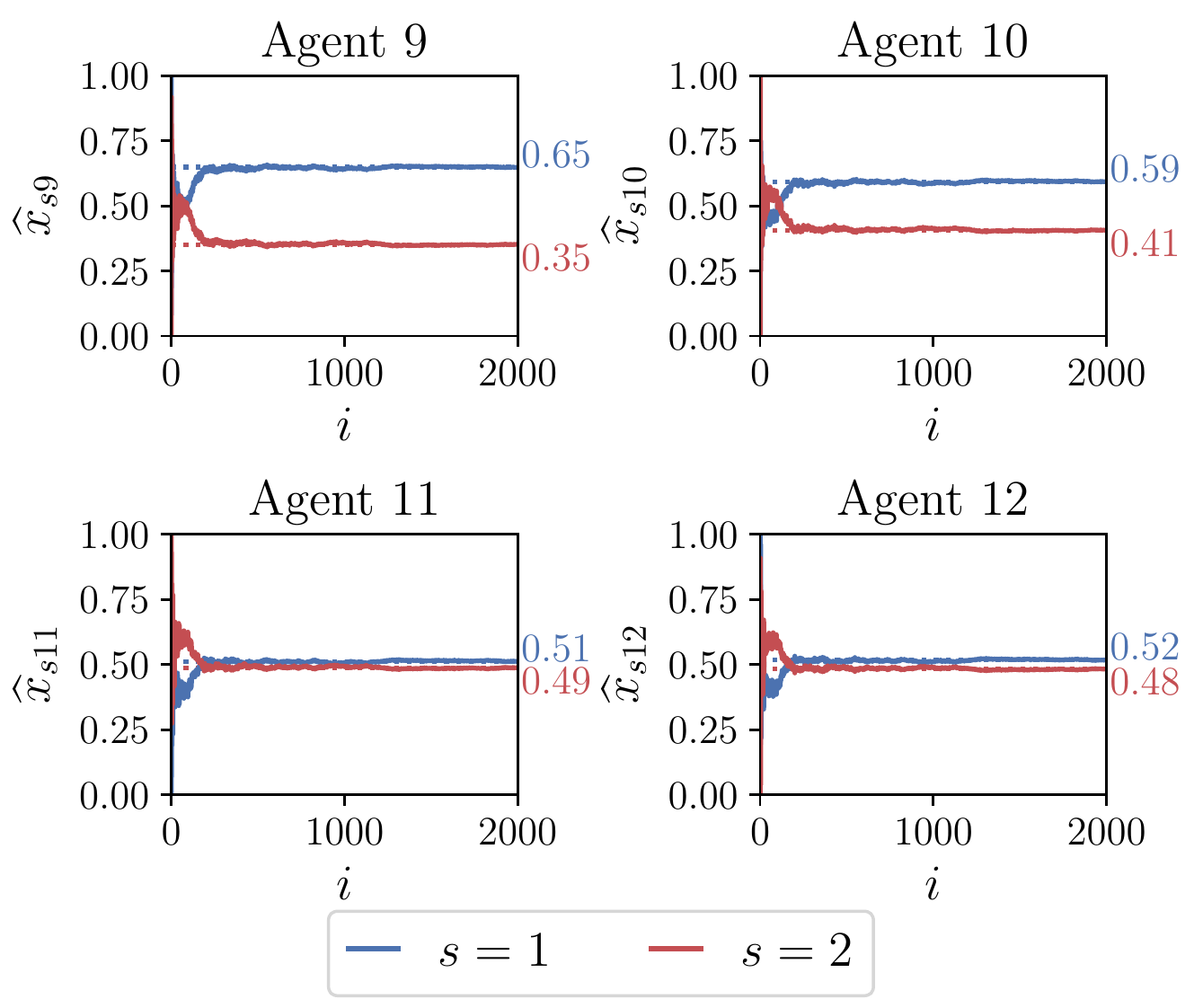}}
\end{minipage}
\caption{
Unperturbed Gaussian model. 
{\em Left}. Network topology. {\em Middle}. Belief convergence at the receiving agents. {\em Right}. Estimated macroscopic topology. For each of the four panels, the numbers on the right denote the true values $\{x_{sk}\}$, with different colors denoting different $s$, according to the legend.}
\label{fig:simTheorem2}
\end{figure}

{\em a) Gaussian with} $H=S=2$.
We consider the topology shown in the leftmost panel of Fig.~\ref{fig:simTheorem2}. The likelihoods and true distributions for the sending sub-networks are unit-variance Gaussian with means $\nu_1=\mathsf{m}_1=1$, $\nu_2=\mathsf{m}_2=2$. The receiving agents\footnote{We recall that the models of the receiving agents will be ultimately immaterial as regards their limiting beliefs.} employ the same likelihoods of the sending agents, and their true distributions are unit-variance Gaussian with mean equal to $1$. In Fig.~\ref{fig:simTheorem2} (middle) we show the convergence of the receiving agents' beliefs.

Next, we address the topology learning problem. 
First, for an observation time $i$, we construct the empirical data $\widehat{y}_k(\theta)=(1/i)\log \bm{\psi}_{k,i}(\theta)$, and construct an estimate $\widehat{\theta}^{\star}_k$ as the value of $\theta$ that maximizes $\widehat{y}_k(\theta)$ (i.e., the hypothesis where $\widehat{y}_k(\theta)$ will collapse to $1$). We can then construct an estimate for $B_k$ as:
\beq
\widehat{B}_k= \left(\mathbbm{1}_{H} e\T_{\widehat{\theta}^{\star}_k} - I_H\right) D,
\eeq
from which we obtain $\widehat{C}_k$ by adding an all-ones row, according to~(\ref{eq:augmentedB}).
At this point, we have verified on the simulated data that, for any receiving agent $k\in\{9,10,11,12\}$, the matrices $\widehat{C}_k$ are full column rank. Then, we used~(\ref{eq:systinvert}) with empirical matrices replacing the exact ones to estimate the connection-weight vector $x_{k}$ as:\footnote{The symbol $\widehat{\phantom{x}}$ is used for quantities {\em estimated} from the data, to be not confused with the symbol $\widetilde{\phantom{x}}$ used for the {\em exact} quantities appearing in~(\ref{eq:linsystopology}).}
%``~
\beq
\widehat{x}_{k}=\widehat{C}_k^{\dagger}\begin{bmatrix}\widehat{y}_k\\1\end{bmatrix}= 
(\widehat{C}_k\T \widehat{C}_k)^{-1}\widehat{C}_k\T \begin{bmatrix}\widehat{y}_k\\1\end{bmatrix}.
\eeq
We see from Fig.~\ref{fig:simTheorem2} (right) that this procedure allows us to retrieve the topology coefficients $\{x_{sk}\}$, provided that the system evolves for a sufficiently long time.

\begin{figure}[t]
\begin{minipage}{.33\linewidth}
\vspace*{-20pt}
{\centering{\bf ~~~~~~network graph}\par\medskip}
%\hspace*{-20pt}
\centering
%\subfloat[]
{
\includegraphics[scale=\paneldim]{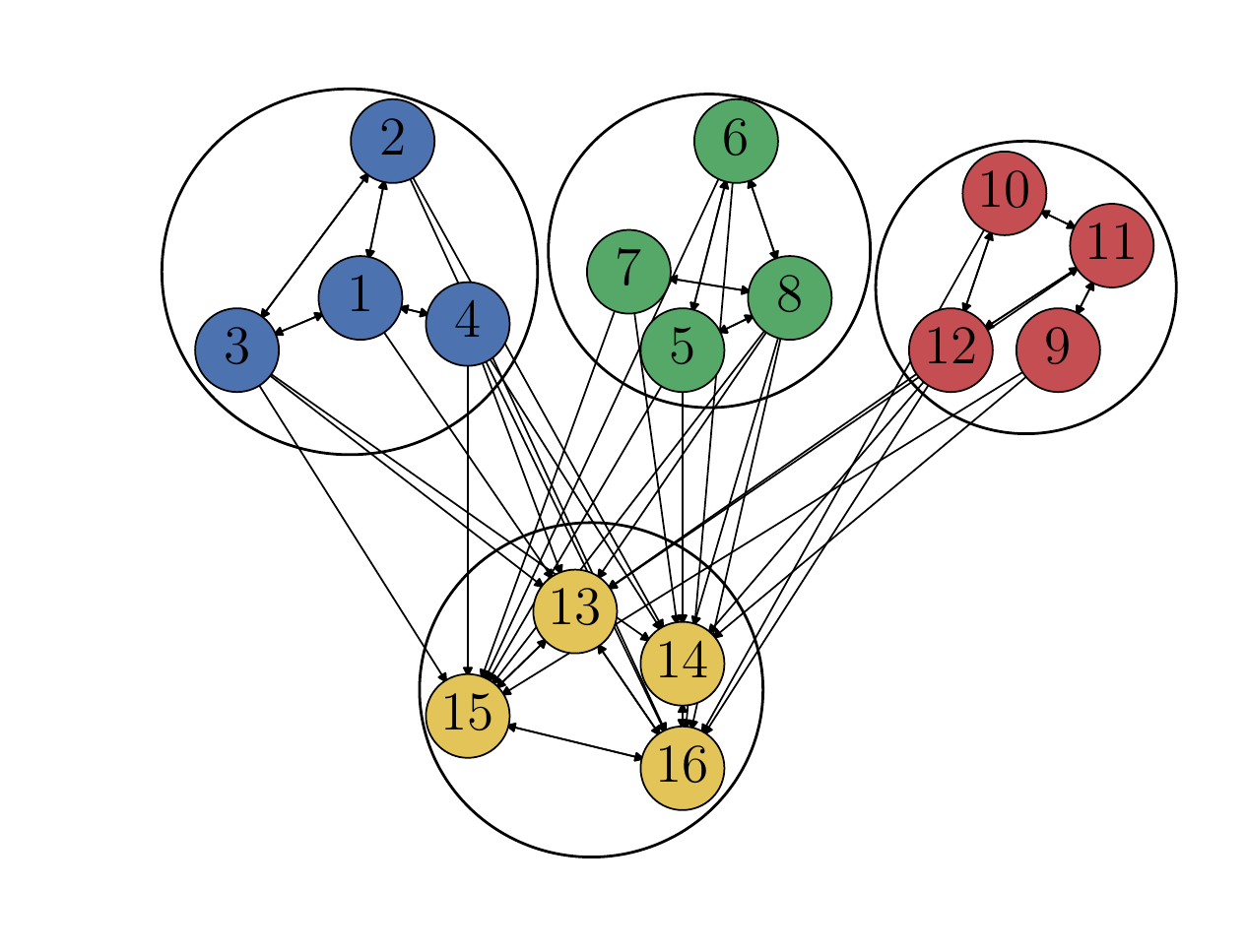}}
\end{minipage}
\begin{minipage}{.33\linewidth}
{\centering{\bf ~~~~~~belief evolution}\par\medskip}
%\vspace*{5pt}
%\hspace*{-10pt}
\centering
%\subfloat[]
{
\includegraphics[scale=\paneldim]{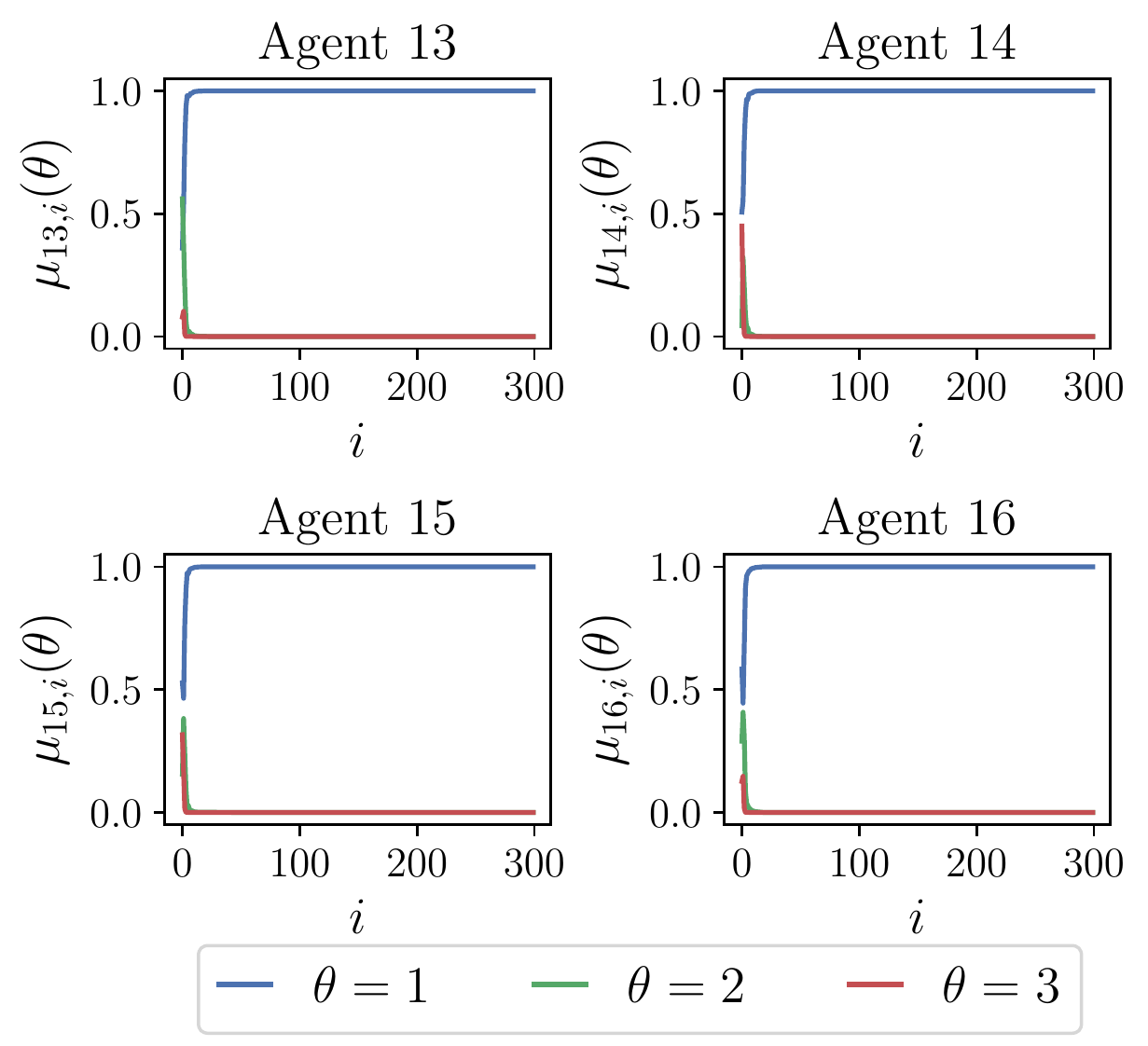}}
\end{minipage}
\begin{minipage}{.33\linewidth}
{\centering{\bf ~~~~~~estimated weights}\par\medskip}
%\vspace*{5pt}
\centering
%\subfloat[]
{
\includegraphics[scale=\paneldim]{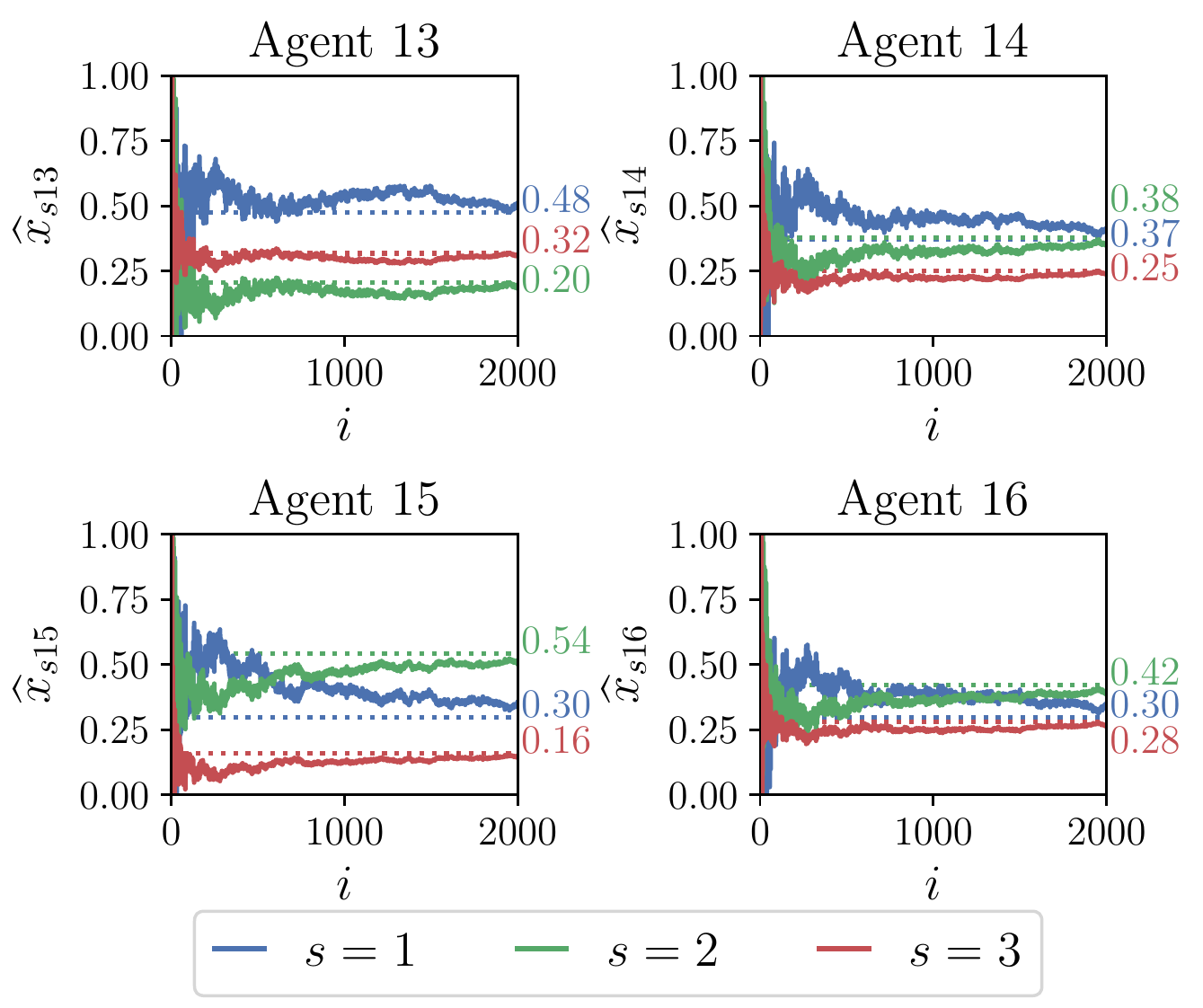}}
\end{minipage}
\caption{
Perturbed Gaussian model. 
{\em Left}. Network topology. {\em Middle}. Belief convergence at the receiving agents. {\em Right}. Estimated macroscopic topology. For each of the four panels, the numbers on the right denote the true values $\{x_{sk}\}$, with different colors denoting different $s$, according to the legend.
}
\label{fig:simTheorem3Gauss}
\end{figure}

{\em b) Randomly perturbed Gaussian with $H=S=3$.}
The network topology has three sending sub-networks and one receiving sub-network as shown in the leftmost panel of Fig.~\ref{fig:simTheorem3Gauss}. 
When $S>2$, we know from Theorem~\ref{theor:TopologyGaussian} that for the {\em structured} Gaussian model, diversity in the sending components is not enough to ensure the full column rank of the matrix $C_k$. 
In order to increase diversity, we consider a {\em randomly perturbed} model for the likelihood functions, where the likelihood of the $s$-th sending sub-network, evaluated at hypothesis $\theta$, is unit-variance Gaussian with mean $\theta+\bm{\epsilon}_{\theta s}$. 
The random variables $\{\bm{\epsilon}_{\theta s}\}$ are equally correlated zero-mean Gaussian with variance equal to $0.02$ and Pearson correlation coefficient equal to $0.5$. 
For the receiving sub-network we use the same type of random perturbation of the likelihoods.
The true distributions for {\em all} sending and receiving agents are unit-variance Gaussian with mean equal to $1$.
%, while the true distributions are unit-variance Gaussian with mean equal to $3$.
%independent across the sending sub-networks. To add %generality, we generated them as correlated across the index $\theta=1,2,\ldots,H$.
%is a random variable sampled from a uniform distribution with support $[-0.1, 0.1]$.
%These expectations are sampled independently across the hypotheses (i.e., w.r.t. $\theta$) and the sending components (i.e., w.r.t. $s$). 
%\beq
%f^{(s)}(\xi)=\frac{1}{\sqrt{2\pi}}\exp\left\{-\frac{(\xi-s)^2}{2}\right\}.
%\eeq
The belief convergence for the receiving agents can be seen in the middle group of panels of Fig.~\ref{fig:simTheorem3Gauss}. 
In the rightmost group of panels, we see how the estimates $\{\widehat{x}_{sk}\}$ of the topology weights converge to the true values $\{x_{sk}\}$.
In contrast with the {\em structured} Gaussian case, topology learning is now feasible for $S>2$ and even if the true distributions are equal across all sending components. 
This change in behavior is due to the {\em diversity} in the models of the sending sub-networks, represented by the different means of the likelihoods. Moreover, we see from the parameters of the random variables $\{\bm{\epsilon}_{\theta s}\}$ that a relatively small perturbation is already sufficient to enable consistent topology learning.

\begin{figure}[t]
\begin{minipage}{.33\linewidth}
\vspace*{-20pt}
{\centering{\bf ~~~~~~network graph}\par\medskip}
%\hspace*{-20pt}
\centering
%\subfloat[]
{
\includegraphics[scale=\paneldim]{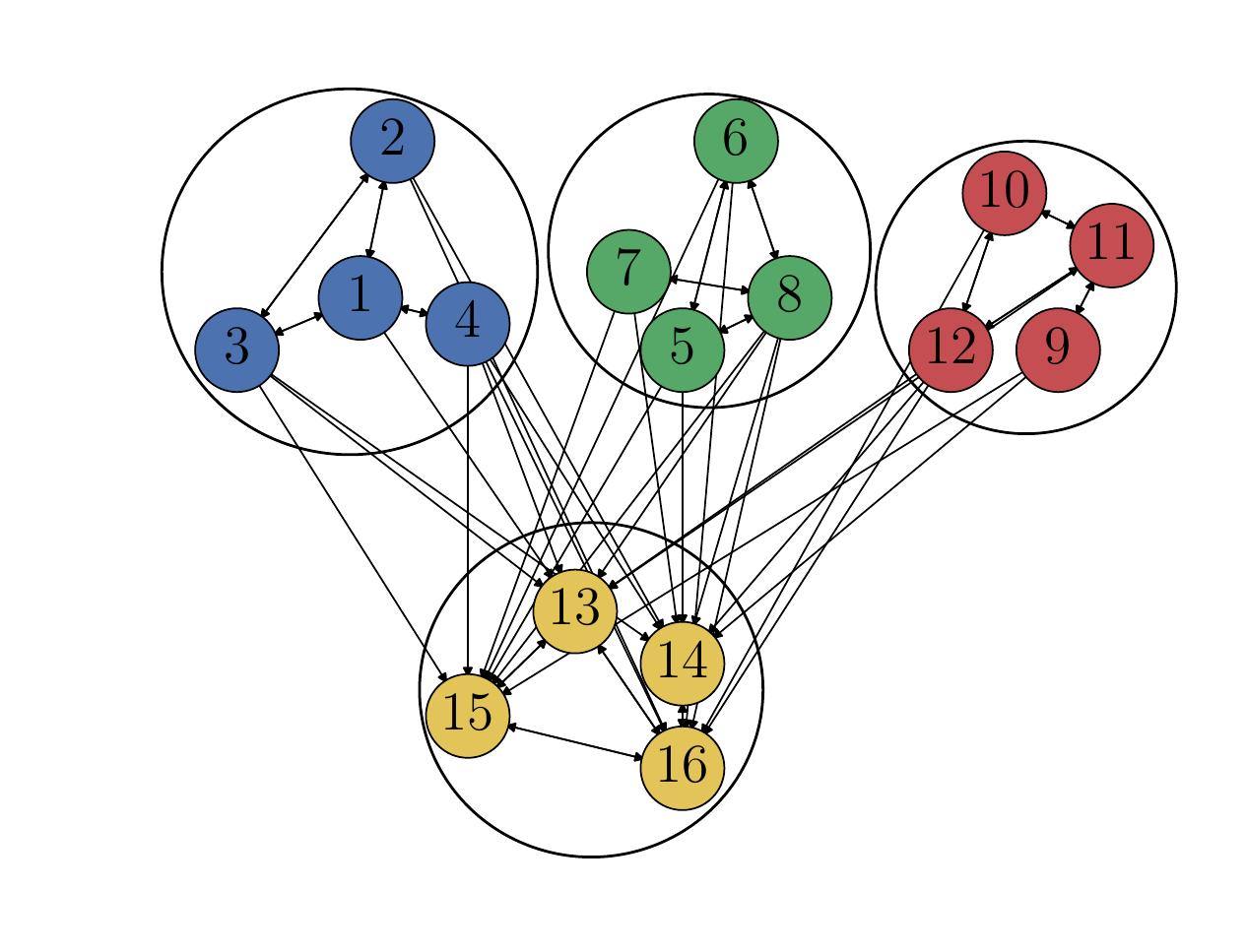}}
\end{minipage}
\begin{minipage}{.33\linewidth}
{\centering{\bf ~~~~~~belief evolution}\par\medskip}
%\vspace*{5pt}
%\hspace*{-10pt}
\centering
%\subfloat[]
{
\includegraphics[scale=\paneldim]{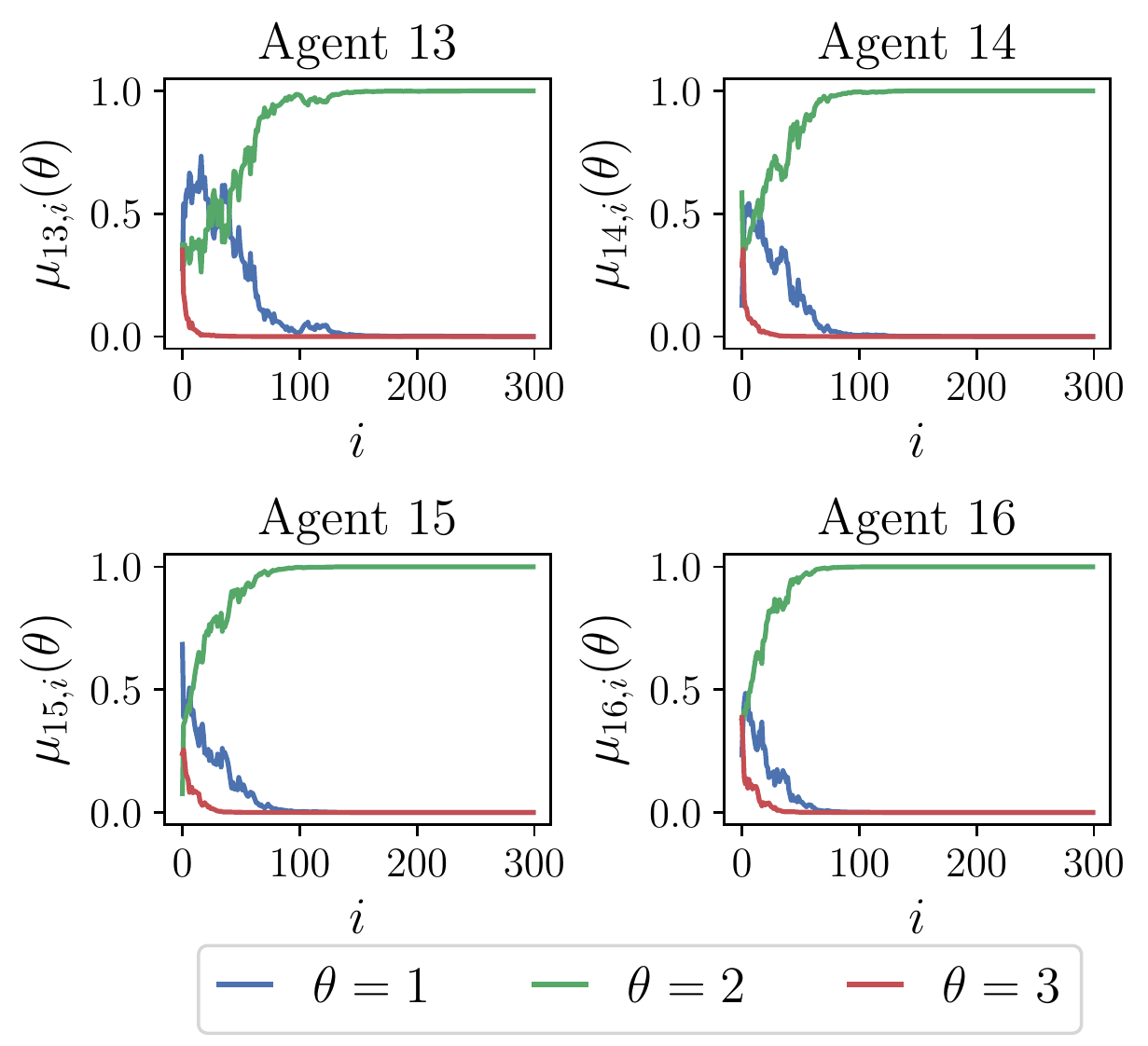}}
\end{minipage}
\begin{minipage}{.33\linewidth}
{\centering{\bf ~~~~~~estimated weights}\par\medskip}
%\vspace*{5pt}
\centering
%\subfloat[]
{
\includegraphics[scale=\paneldim]{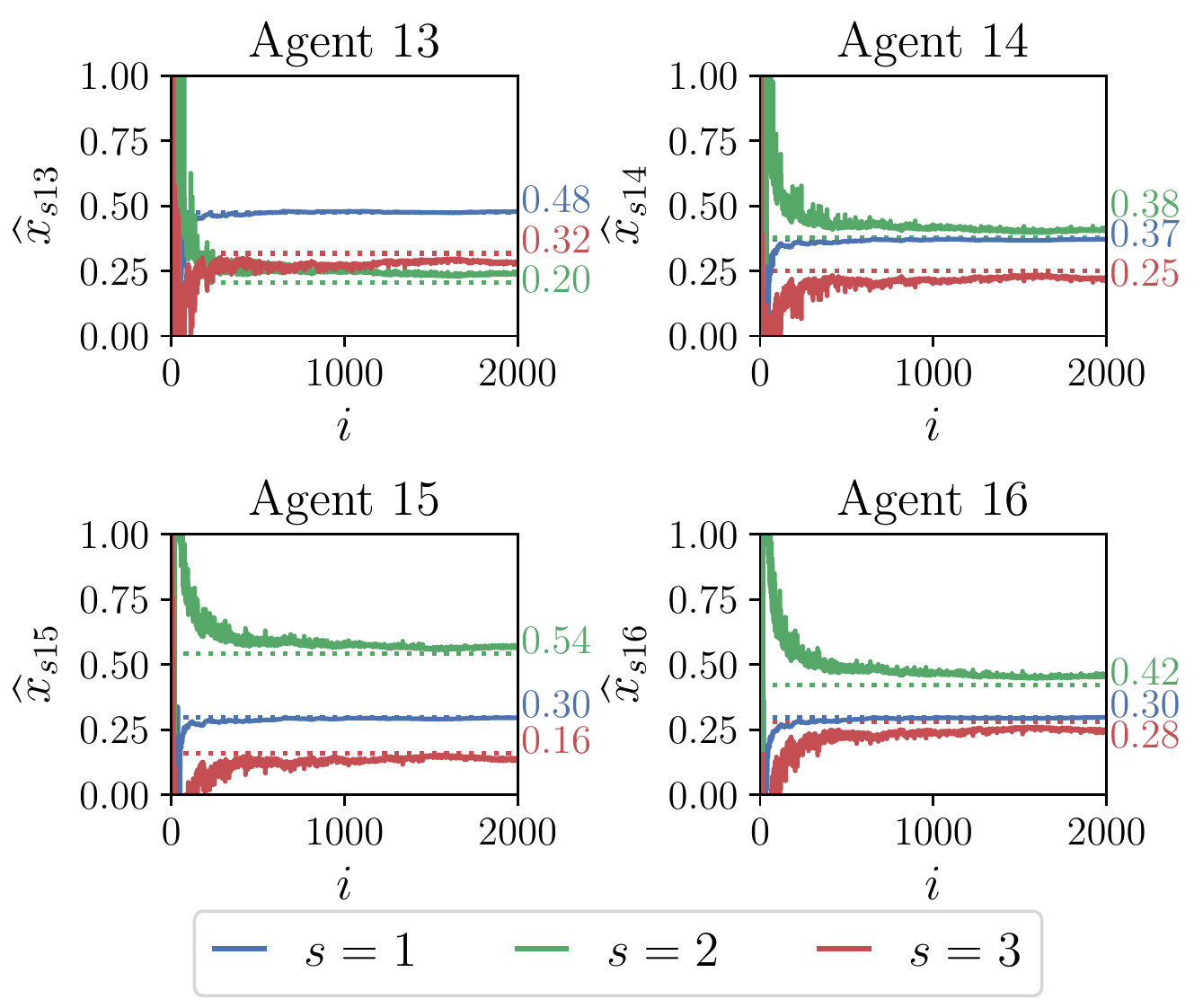}}
\end{minipage}
\caption{Perturbed Beta model. 
{\em Left}. Network topology. {\em Middle}. Belief convergence at the receiving agents. {\em Right}. Estimated macroscopic topology. For each of the four panels, the numbers on the right denote the true values $\{x_{sk}\}$, with different colors denoting different $s$, according to the legend.}
\label{fig:simTheorem3Beta}
\end{figure}

{\em c) Beta with} $H=S=3$.
Finally, we consider a non-Gaussian example. 
The network topology is the same as in the last example.
However, now the likelihood functions follow a Beta distribution with scale parameter equal to $2$ and with shape parameters given by $\theta + 1+\bm{u}_{\theta s}$, where $\{\bm{u}_{\theta s}\}$, for $\theta\in\{1,2,3\}$ and $s\in\{1,2,3\}$,  are independent random variables sampled from a uniform distribution with support $[-0.1, 0.1]$. 
The true distributions coincide with the unperturbed likelihoods, i.e., the true distribution of the $s$-th sending sub-network is a Beta distribution with scale parameter equal to $2$ and shape parameter equal to $s+1$. 
For the receiving sub-network we apply the same type of random perturbation of the likelihoods, whereas the true distributions are Beta with scale and shape parameters equal to $2$. The belief convergence for the receiving agents can be seen in the middle group of panels of Fig.~\ref{fig:simTheorem3Beta}. 
In the rightmost group of panels, we see the convergence of the topology estimates.

\section{Social Learning vs. Topology Learning}
\label{sec:SLvsTL}
In this work we have considered two learning problems. The first problem is the Social Learning (SL) problem, which is the goal of the 
agents in the network. These agents aim at forming their opinions after consulting the beliefs of their neighbors through an iterative update-and-combine SL algorithm.
The second problem is the Topology Learning (TL) problem, where a receiving agent (or some entity monitoring its behavior) attempts to get knowledge about the connections between that receiving agent and the sending sub-networks.
We can refer to the SL problem as the {\em direct} learning problem, in the sense that it is the original inferential problem the network is deployed for.
Likewise, we can refer to the TL problem as the {\em dual} learning problem, since it is an inferential procedure that takes as input data {\em the output of the direct TL  problem}. 

The analysis conducted in this work has revealed some interesting interplay between SL and TL problems. 
%The feasibility of the TL problem for the different settings addressed in this work 
%is represented in Table~\ref{tab:???}. 
Let us make a summary of the main results. We recall that $S$ denotes the number of sending sub-networks, and $H$ the number of hypotheses.
First, we established in Lemma~\ref{prop:necessary} that $H\geq S$ is a necessary condition to achieve consistent TL. 
This condition has a remarkable interpretation. In a sense, the number of hypotheses is an index (even if not the only one) of complexity associated to the SL problem since, other conditions being equal, more hypotheses make the SL problem more complicated.
Likewise, the number of sending components represents an index of complexity of the TL problem, 
since, other conditions being equal,  estimating more links is more complicated.   
According to these remarks, the condition $H \geq S$ {\em implies that the TL problem can be feasible when its complexity 
is not greater than the complexity of the SL problem}. Such an interplay appears to be not obvious at all. 
As a matter of fact, in the traditional topology inference problems, the connections between agents are inferred from some kind of pairwise measure of their dependency. In our setting, since we cannot measure the output of the 
sending sub-network, we cannot get direct data quantifying dependency between a receiving and a sending agent. 
Our TL inference is based instead on the belief functions. 
The belief function contains some richness of information in that it is evaluated for the $H$ different values of $\theta$. 
This richness (i.e., $H$) is critical to enable feasibility of the TL problem. 
In particular, $H\geq S$ means that the richness of information in the belief function should be greater than or equal to the number of unknown topology weights to be estimated, $S$.

Having established a necessary condition for consistent TL, we moved on to examine some useful models to see whether and when consistent TL is in fact achievable.
First, we have considered a structured Gaussian model where {\em all} sending sub-networks use the same family of Gaussian likelihoods, and the sending sub-networks have distinct true distributions, each one coinciding with one of the likelihoods. 
We have shown in Theorem~\ref{theor:TopologyGaussian} that the TL problem is feasible only if $S=2$, for any $H\geq 2$.
%On the other hand, when the number of sending components is greater than $2$, 
%then the TL problem is not feasible.
The limited possibility of achieving consistent TL can be ascribed to the limited diversity existing between the different sub-networks (which all use the same family of likelihoods). 
%This lack of diversity turns out to be detrimental for the TL problem. 
%similarity induces limited diversity across the sending sub-networks, and a reduced degree of diversity in the statistical models is 
This observation motivated the analysis of more general models with a certain degree of {\em diversity}, a condition formalized by saying that the KL divergences between true distributions and likelihoods are not structured, i.e., they are nonnegative real numbers with no particular relationship among them.
Under this setting we have ascertained that, if $H\geq S$, the TL problem becomes feasible for almost all configurations, in a precise mathematical sense as stated in Theorem~\ref{theor:TopologyRandom}. 
%with probability one whenever $H\geq S$. 
In summary, two critical features that enable consistent TL are: {\em more hypotheses than sending components} and {\em a sufficient degree of diversity}.
%However, we have also shown that in the structured Gaussian model the TL problem is not feasible for $S>2$. Why? 
%In contrast, the TL problem becomes feasible when {\em sufficient 
%diversity exists}, which is the main conclusion revealed by Theorem~\ref{theor:TopologyRandom}.

\begin{appendices}

\section{Proof of Theorem~\ref{theor:limbelief}}
\label{app:proofTheor1}
Exploiting~(\ref{eq:private}) we can write, for $\theta,\theta^{\prime}\in\Theta$:
\beq
\log \frac{\bm{\psi}_{\ell,i}(\theta)}{\bm{\psi}_{\ell,i}(\theta^{\prime})}
=
\log\frac
{\bm{\mu}_{\ell,i-1}(\theta)}
{\bm{\mu}_{\ell,i-1}(\theta^{\prime})}
+
\log\frac
{L_\ell(\bm{\xi}_{\ell,i} | \theta)}
{L_\ell(\bm{\xi}_{\ell,i} | \theta^{\prime})}.
\label{eq:psi}
\eeq
Using~\eqref{eq:combine} we have:
\beq
\log\frac{\bm{\mu}_{k,i}(\theta)}{\bm{\mu}_{k,i}(\theta^{\prime})}=
\sum_{\ell=1}^{N}a_{\ell k}
\left(\log\frac
{\bm{\mu}_{\ell,i-1}(\theta)}
{\bm{\mu}_{\ell,i-1}(\theta^{\prime})}
+
\log\frac
{L_\ell(\bm{\xi}_{\ell,i} | \theta)}
{L_\ell(\bm{\xi}_{\ell,i} | \theta^{\prime})}\right).
\label{eq:firstbeliefremastering}
\eeq
By iterating over $i$, we can write:
\beqa
\frac{1}{i}\log\frac{\bm{\mu}_{k,i}(\theta)}{\bm{\mu}_{k,i}(\theta^{\prime})}&=&
\frac{1}{i}
\sum_{\ell=1}^N
\sum_{t=1}^i
[A^{i - t +1}]_{\ell k}
\displaystyle{
\log \frac{L_{\ell}(\bm{\xi}_{\ell,t}|\theta)}{L_{\ell}(\bm{\xi}_{\ell,t}|\theta^{\prime})}
}
\nonumber\\
&+& \frac{1}{i}
\sum_{\ell=1}^N[A^i]_{\ell k}
\log\frac{\bm{\mu}_{\ell,0}(\theta)}{\bm{\mu}_{\ell,0}(\theta^{\prime})}.
\label{eq:fundexpan}
\eeqa
Under Assumptions~\ref{assum:divergences}--\ref{assum:initbel}, thanks to the integrability of the log-ratios between the true distributions and the likelihoods implied by~(\ref{eq:finiteKLdiv}), through standard limiting arguments (see, e.g.,~\cite{Javidi,NedicTAC2017}) it is possible to determine the asymptotic behavior of~(\ref{eq:fundexpan}) by: $i)$ replacing the powers of matrix $A$ with their limit $A_{\infty}$ in~(\ref{eq:limatweak}); and $ii)$ applying the strong law of large numbers to conclude that:
\beqa
\lefteqn{\lim_{i\rightarrow\infty}
\frac{1}{i}\log\frac{\bm{\mu}_{k,i}(\theta)}{\bm{\mu}_{k,i}(\theta^{\prime})}
}
\nonumber\\
&\stackrel{\textnormal{a.s.}}{=}&
\sum_{\ell=1}^N
[A_{\infty}]_{\ell k}
\E\left[
\log \frac{f_{\ell}(\bm{\xi}_{\ell,t})}{L_{\ell}(\bm{\xi}_{\ell,t}|\theta^{\prime})}
-
\log \frac{f_{\ell}(\bm{\xi}_{\ell,t})}{L_{\ell}(\bm{\xi}_{\ell,t}|\theta)}
\right]\nonumber\\
&=&
\sum_{\ell\in\mathcal{S}}
\omega_{\ell k} D[f_{\ell}||L_{\ell}(\theta^{\prime})]
-
\sum_{\ell\in\mathcal{S}}
\omega_{\ell k} D[f_{\ell}||L_{\ell}(\theta)]\nonumber\\
&=&
\mathscr{D}_{k}(\theta^{\prime})-\mathscr{D}_{k}(\theta),
\label{eq:logratioconv}
\eeqa
where in the second-last equality we used~(\ref{eq:KLdivfirst}), and we performed the replacement $[A_{\infty}]_{\ell k}=\omega_{\ell k}$, which holds in view of the block representation in~(\ref{eq:limatweak}) since $k$ is a receiving agent and since we adopt the indexing in~(\ref{eq:recindex}).
Now, in light of Assumption~\ref{assum:uniquemin}, we conclude that:
\beq
\lim_{i\rightarrow\infty}
\frac{1}{i}\log\frac{\bm{\mu}_{k,i}(\theta)}{\bm{\mu}_{k,i}(\theta^{\star}_k)}
\stackrel{\textnormal{a.s.}}{=}
\mathscr{D}_{k}(\theta^{\star}_k)-\mathscr{D}_{k}(\theta)<0
\label{eq:negexpon}
\eeq
for all $\theta\neq\theta^{\star}_k$. 
Since the denominator of $\frac{\bm{\mu}_{k,i}(\theta)}{\bm{\mu}_{k,i}(\theta^{\star}_k)}$ is bounded by $1$, Eq.~(\ref{eq:negexpon}) implies that the numerator $\bm{\mu}_{k,i}(\theta)$ is converging to zero. Since the belief function must sum to $1$, the result in~(\ref{eq:limbelief1st}) holds.

\section{Proof of Theorem~\ref{theor:TopologyGaussian}}
\label{app:ProofTheor2}
Preliminarily, it is useful to introduce some auxiliary matrices. 
We let, for all $\theta=1,2,\ldots,H$:
\beq
\mathcal{I}(\theta)\dfz\mathbbm{1}_{H}e_{\theta}\T-I_H,
\label{eq:matIdef}
\eeq
and
\beq
B(\theta)\dfz\mathcal{I}(\theta)D,\quad
C(\theta)=
\begin{bmatrix}
B(\theta)\\
\mathbbm{1}_S\T
\end{bmatrix}.
\label{eq:auxiliarymat}
\eeq
In view of Eqs.~(\ref{eq:Bkevec}) and~(\ref{eq:augmentedB}), the definitions in~(\ref{eq:matIdef}) and~(\ref{eq:auxiliarymat}) imply:
\beq
B_k=B(\theta^{\star}_k),\quad C_k=C(\theta^{\star}_k).
\label{eq:BkCk}
\eeq
We continue by showing some useful properties of the matrix $D$ under the considered Gaussian model.
Let us focus on the representation in~(\ref{eq:generalD}).
It is a known result that the rank of a Euclidean distance matrix with $n$ points in $\mathbb{R}^{{\sf dim}}$ is at most ${\sf dim}+2$~\cite{dokmanic2015euclidean}. 
Since in our case ${\sf dim}=1$, we, can write:
\beq
\mathrm{rank}(E_S)\leq 3.
\label{eq:EDMrankbound}
\eeq
Moreover, for the cases $S=2$ and $S=3$ we have that:
\begin{align}
	E_2=&\frac 1 2\begin{bmatrix}
	0&(\mathsf{m}_1-\mathsf{m}_2)^2\\(\mathsf{m}_2-\mathsf{m}_1)^2&0
	\end{bmatrix},\\
	E_3=&\frac 1 2\begin{bmatrix}
	0&(\mathsf{m}_1-\mathsf{m}_2)^2&(\mathsf{m}_1-\mathsf{m}_3)^2\\(\mathsf{m}_2-\mathsf{m}_1)^2&0&(\mathsf{m}_2-\mathsf{m}_3)^2\\(\mathsf{m}_3-\mathsf{m}_1)^2&(\mathsf{m}_3-\mathsf{m}_2)^2&0
	\end{bmatrix},
\end{align}
and, hence:
\beqa
\mathrm{det}(E_2)&=&-\frac 1 4(\mathsf{m}_1-\mathsf{m}_2)^2,\nonumber\\
\mathrm{det}(E_3)&=&\frac 1 4\,(\mathsf{m}_1-\mathsf{m}_2)^2 (\mathsf{m}_1-\mathsf{m}_3)^2 (\mathsf{m}_2-\mathsf{m}_3)^2.\nonumber\\
\eeqa
Therefore, when the points that determine the Euclidean distance matrix are all distinct, both the above matrices are full rank.
Thus, when $S=2$, we have that $\mathrm{rank}(E_S)=2$. 
When $S>2$, since $E_3$ is full rank, and in view of~(\ref{eq:EDMrankbound}), we have instead $\mathrm{rank}(E_S)=3$. 
From the representation of $D$ in~(\ref{eq:generalD}), we then conclude that:
\beq
\mathrm{rank}(D)=\left\{
\begin{array}{ll}
2,\qquad \textnormal{if $S=2$},\\
3,\qquad \textnormal{if $S>2$}.
\end{array}
\right.
\label{eq:rankDequal}
\eeq
Next we state and proof a useful lemma.
\begin{lemma}
\label{lem:propertyofI}
Let $\mathcal{I}(\theta)$ be defined as in~(\ref{eq:matIdef}).
Then, for all $\theta=1,2,\ldots,H$ we have that:
\beq
\boxed{
I_H-\mathcal{I}^{\dagger}(\theta)\mathcal{I}(\theta)=\frac{1}{H}\mathbbm{1}\mathbbm{1}\T
}
\label{eq:5}
\eeq
\end{lemma}
\begin{IEEEproof}[Proof of Lemma~\ref{lem:propertyofI}]
For ease of notation, in the following proof the explicit dependence on $\theta$ is suppressed, and we write $\mathcal{I}$ in place of $\mathcal{I}(\theta)$.
By definition of the Moore-€"Penrose inverse, matrix $\mathcal{I}^\dagger$ satisfies: 
\beq
	\mathcal{I}\mathcal{I}^\dagger \mathcal{I}=\mathcal{I},\qquad
	(\mathcal{I}^{\dagger}\mathcal{I})\T= \mathcal{I}^{\dagger}\mathcal{I}.
	 \label{eq:7}
\eeq
Then we note that:
\beq
\mathcal{I}(I_H-\mathcal{I}^{\dagger}\mathcal{I})=\mathcal{I}- \mathcal{I}\mathcal{I}^\dagger \mathcal{I}
=\mathcal{I}-\mathcal{I}=0, 
\label{eq:6}
\eeq
where in the second equality we used the first identity in~\eqref{eq:7}. 
Equation~(\ref{eq:6}) implies that the columns of $(I_H-\mathcal{I}^{\dagger}\mathcal{I})$ belong to the null space of $\mathcal{I}$, denoted by $\mathscr{N}(\mathcal{I})=\{v: \mathcal{I} v=0 \}$. 
On the other hand, in view of~(\ref{eq:matIdef}) we can write:
\beq
\mathcal{I}v = \mathbbm{1}_H e\T_{\theta} v-v= \mathbbm{1}_H v_{\theta}-v=0,
\label{eq:11}
\eeq
with $v_{\theta}$ the $\theta$-th element of $v$. As a result, Eq.~(\ref{eq:11}) will be satisfied only if $v_h = v_{\theta}$ for all $h=1,\dots, H$. Therefore, we obtain:
\beq
\mathscr{N}(\mathcal{I})=\{\alpha \mathbbm{1}_H: \alpha \in \mathbb{R} \},
\eeq
further implying, in light of~(\ref{eq:6}), that, for each $h=1,2,\ldots,H$, the $h$-th column of $I_H-\mathcal{I}^\dagger\mathcal{I}$ is of form $\alpha_h\mathbbm{1}_H$ for some $\{\alpha_h\}$. 
On the other hand, since $I_H-\mathcal{I}^\dagger\mathcal{I}$ is symmetric
in view of the second identity in \eqref{eq:7}, we conclude that $\alpha_h=\overline{\alpha}$ for all $h$, namely,
\beq
I_H-\mathcal{I}^{\dagger}\mathcal{I} = \overline{\alpha} \mathbbm{1}_H \mathbbm{1}_H\T
\label{eq:alphabar}
\eeq
for some $\overline{\alpha} \in \mathbb{R}$. Finally, since in particular $\mathbbm{1}_H\in\mathscr{N}(\mathcal{I})$, we can write:
\beq
(I_H-\mathcal{I}^{\dagger}\mathcal{I} )\mathbbm{1}_H=
\mathbbm{1}_H-\mathcal{I}^{\dagger}\mathcal{I} \mathbbm{1}_H = \mathbbm{1}_H,
\eeq
which, in view of~(\ref{eq:alphabar}), yields: 
\beq
\overline{\alpha} \,\mathbbm{1}_H \mathbbm{1}_H\T\mathbbm{1}_H = 
\overline{\alpha} H \mathbbm{1}_{H}=\mathbbm{1}_H\Rightarrow \overline{\alpha}=\frac 1 H, 
\eeq
and we have in fact proved \eqref{eq:5}.
\end{IEEEproof}

We are now ready to prove Theorem~\ref{theor:TopologyGaussian}.

\begin{IEEEproof}[Proof of Theorem~\ref{theor:TopologyGaussian}]
We will now show that
\beq
\mathrm{rank}(C(\theta))=2~~~\textnormal{for all}~\theta=1,2,\ldots,H,
\eeq
which clearly implies the claim of the theorem in view of the second equation in~(\ref{eq:BkCk}).

For the case $H=S=2$, it is immediately seen that the matrix $C(\theta)$ (assuming, e.g., $\theta=1$) takes on the form:
\beq
C(1)=\frac 1 2
\begin{bmatrix}
0&0\\
0&d_{12}
\\
1&1
\end{bmatrix},
\eeq
which reveals that $\mathrm{rank}(C(\theta))=2$.

Let us move on to examine the other cases where $H\geq S$ (excluding $H=S=2$). We will examine first the properties of the matrix $B(\theta)$ in~(\ref{eq:auxiliarymat}). As done before, the dependence on $\theta$ is suppressed for ease of notation, and, in particular, we write $B$, $C$, and $\mathcal{I}$ in place of $B(\theta)$, $C(\theta)$, and $\mathcal{I}(\theta)$, respectively. 
Applying Sylvester's inequality to the first equation in~(\ref{eq:auxiliarymat}) we can write~\cite{horn2012matrix}:
\beq
	\mathrm{rank}(B)\geq \mathrm{rank}(D)+\mathrm{rank}(\mathcal{I})-H=\mathrm{rank}(D)-1,
	\label{eq:1}
\eeq
where in the latter equality we used the fact that $\mathrm{rank}(\mathcal{I})=H-1$.
Therefore, from~(\ref{eq:rankDequal}) and~(\ref{eq:1}) we conclude that:
\begin{align}
	&\mathrm{rank}(B)\geq 1, ~~~~ \text{if }S=2 \label{eq:2}\\
	&\mathrm{rank}(B)\geq 2, ~~~~ \text{if }S>2. \label{eq:3}
\end{align}
Now we would like to see if equality is satisfied for the cases $S=2$ (with $H>2$) and $S>2$ (with $H\geq S$). 

To this end, we start by noticing that equality in Sylvester's inequality holds if, and only if, there exist matrices $X$ and $Y$ that solve~\cite{horn2012matrix}:
\beq
D X+Y\mathcal{I}=I_H,
\label{eq:mainmatrixeq}
\eeq
which in turn admits a solution if, and only if,~\cite{baksalary1979matrix}:
\beq
(I_H-DD^\dagger)(I_H-\mathcal{I}^\dagger \mathcal{I})=0.
\label{eq:4}
\eeq
Applying Lemma~\ref{lem:propertyofI}, from \eqref{eq:4} we get:
\beq
(I_H-D D^\dagger)\frac{1}{H}\mathbbm{1}_H\mathbbm{1}_H\T=0
\eeq
which means that the equality sign in~(\ref{eq:2}) or~(\ref{eq:3}) holds if, and only if:
\beq
\boxed{
DD^\dagger \mathbbm{1}_H=\mathbbm{1}_H
}
\label{eq:10}
\eeq
In particular, we will now show that \eqref{eq:10} does not hold for $S=2$, while it holds for $S>2$. 

Let us start with the case $S=2$ (and $H>2$). 
We will appeal to the representation of $D$ in~(\ref{eq:generalD}), which for the case $S=2$ can be written as:
\beq
D=\frac 1 2
\begin{bmatrix}
0&(\mathsf{m}_1-\mathsf{m}_2)^2\\
(\mathsf{m}_2-\mathsf{m}_1)^2&0\\
(\mathsf{m}_3-\mathsf{m}_1)^2&(\mathsf{m}_3-\mathsf{m}_2)^2\\
\vdots&\vdots\\
(\mathsf{m}_H-\mathsf{m}_1)^2&(\mathsf{m}_H-\mathsf{m}_2)^2\\
\end{bmatrix}.
\eeq
Let us now consider the linear system $D v=\mathbbm{1}_H$. From the first two rows of $D$, we get the unique solution:
$v=2(\mathsf{m}_1-\mathsf{m}_2)^{-2}\mathbbm{1}_2$.
Considering now the third row, we get the identity $(\mathsf{m}_3-\mathsf{m}_1)^2+(\mathsf{m}_3-\mathsf{m}_2)^2=(\mathsf{m}_1-\mathsf{m}_2)^2$,
which is true only if the third point, $\mathsf{m}_3$, is equal to one of the previous points. 
We conclude that there exist no $v$ such that $D v=\mathbbm{1}_{H}$, which further implies that
$D D^\dagger  \mathbbm{1}_H\neq  \mathbbm{1}_H$
Therefore, for $S=2$ Eq.~\eqref{eq:2} gives $\mathrm{rank}(B)>1$, which since $B$ is of dimension $H\times2$, with $H>2$, implies that $\mathrm{rank}(B)=2$.

Let us move on to examine the case $S>2$ and $H\geq 2$. 
It is known that, for an $L\times L$ Euclidean distance matrix $M$, one has $M M^{\dagger}\mathbbm{1}_L = \mathbbm{1}_L$, implying that $\mathbbm{1}_L$ belongs to the range space of $M$~\cite{gower1985properties}.
We can apply this result to the matrices $E_S$ and $E_H$ in~(\ref{eq:generalD}), since they are proportional to Euclidean distance matrices. In particular, we can say that there exist vectors $u_S$ and $u_H$ such that $E_S u_S=\mathbbm{1}_S,\qquad E_H u_H=\mathbbm{1}_H$.
In particular, one of the (infinite) solutions is given by
\beq
u^{\star}_{H}=
\begin{bmatrix}
u_{S}\\0
\end{bmatrix}.
\label{eq:uHstar}
\eeq 
Applying now~(\ref{eq:uHstar}) into~(\ref{eq:generalD}), we can write:
\beq
\mathbbm{1}_H=
E_H u^{\star}_H=
\begin{bmatrix}
E_S&F\T\\
F&E_{H-S}
\end{bmatrix}
\begin{bmatrix}
u_{S}\\0
\end{bmatrix}
=
\begin{bmatrix}
E_S\\
F
\end{bmatrix}
u_S=D u_S.
\eeq
Equation~(\ref{eq:10}) now follows by observing that:
\beq
D D^{\dagger}\underbrace{\mathbbm{1}_H}_{D u_S}=\underbrace{D D^{\dagger} D}_{D} u_S=D u_S=\mathbbm{1}_H.
\eeq
We have in fact shown that \eqref{eq:10} holds true for $S>2$, which implies that \eqref{eq:3} becomes an equality for $S>2$. 

In summary, we have shown so far that $\mathrm{rank}(B)=2$ for all $H\geq S$ (but for the case $H=S=2$, which has been examined separately). 
We will now use this result to prove the claim of the theorem, namely, that $\mathrm{rank}(C)=2$. 
Since $C$ is obtained from $B$ by adding an all-ones row, determining the rank of $C$ from that of $B$ amounts to check whether the row vector $\mathbbm{1}_S\T$ lies in the row space of $B$, which is tantamount to ascertaining whether there exists $z$ such that: 
\beq
z\T \mathcal{I}D=\mathbbm{1}\T_S.
\label{eq:4new}
\eeq
Since we exclude the case $H=S=2$, we have always $H\geq 3$. 
Now, let us consider an EDM $E_3$ defined on $3$ distinct points $p_1$, $p_2$, $p_3$. Since in this case $E_3$ is full rank, the system $v_3\T E_3=\mathbbm{1}\T_3$ has the following (unique) solution: 
\beq
v_3\T=
\begin{bmatrix}
\frac{e_{13}+e_{12}-e_{23}}{e_{13}e_{12}}&
\frac{e_{12}+e_{23}-e_{13}}{e_{12}e_{23}}&
\frac{e_{13}+e_{23}-e_{12}}{e_{13}e_{23}}
\end{bmatrix},
\label{eq:v3sol}
\eeq
where we denoted by $e_{i j}=1/2 (p_i-p_j)^2$ the $(i,j)$-th entry of $E_3$. 
Let us now introduce the vector:
\beq
v_H\T=[v_3\T ~~0_{H-3}\T].
\label{eq:vHaug}
\eeq
Since, for $H\geq 3$, we know that $\mathrm{rank}(E_H)=3$, we conclude that:
\beq
v_3\T E_3=\mathbbm{1}_3\T
\Rightarrow
v_H\T E_H=\mathbbm{1}_H\T,
\eeq
which, using the block representation of $D$ in~(\ref{eq:generalD}), yields:
\beq
v_H\T D=\mathbbm{1}_S\T. 
\label{eq:solutD}
\eeq
In view of~(\ref{eq:solutD}), one solution $z$ to~(\ref{eq:4new}) exists if $z\T \mathcal{I}=v_H\T$, that is, if $v_H\T$ lies in the row space of $\mathcal{I}$. 

On the other hand, from the definition in~(\ref{eq:matIdef}), we see that the matrix $\mathcal{I}$ can be represented as:
\beq
\mathcal{I}=
\left[
\begin{array}{ccccccc}
-1 &0&\dots&0&\bm{1}&0\dots 0\\
0 &-1&\dots&0&\bm{1}&0\dots 0\\
\vdots&\vdots&&\vdots\\
0 &0&\dots&-1&\bm{1}&0\dots 0\\
\bm{0} &\bm{0}&\dots&\bm{0}&\bm{0}&\bm{0}\dots \bm{0}\\
0 &0&\dots&0&\bm{1}&-1\dots 0\\
\vdots&\vdots&&\vdots\\
0 &0&\dots&0&\bm{1}&0\dots -1\\
\end{array}
\right],
\label{eq:Imatrepresent}
\eeq
where the bold notation highlights the $\theta$-th row and column. 
According to~(\ref{eq:Imatrepresent}), the row space of $\mathcal{I}$ is: 
\beq
\mathrm{Row}(\mathcal{I})=\left\{
[\alpha_1 ~\alpha_2 \dots \alpha_H]: \alpha_{\theta}=-\sum_{h\neq \theta}\alpha_h
\right\},
\label{eq:rowspaceI}
\eeq
which is equivalent to:
\beq
\mathrm{Row}(\mathcal{I})=\left\{
[\alpha_1 ~\alpha_2 \dots \alpha_H]: ~\alpha\T \mathbbm{1}_H=0\right\}.
\label{eq:rowspaceI}
\eeq
Examining~(\ref{eq:v3sol}), from straightforward algebra it can be shown that $v_3\T\mathbbm{1}_3=0$, which, in light of~(\ref{eq:vHaug}), implies that $v_H\T\mathbbm{1}_H=0$. Using~(\ref{eq:rowspaceI}), we conclude that $v_H\T$ lies in fact in the row space of $\mathcal{I}$, which finally implies, for $H\geq S$ (excluding the case $H=S=2$) that $\mathrm{rank}(C)=2$.
\end{IEEEproof}

\section{Proof of Theorem~\ref{theor:TopologyRandom}}
\label{app:ProofTheor3}
We remark that in our setting the divergences are modeled as random variables, which implies that the value of $\bm{\theta}^{\star}_k$ is random as well. We should take this into account when proving the claim of the theorem. 
First, we observe that:
\beqa
\lefteqn{\P[\bm{\theta}_k^{\star} \textnormal{ is unique and } \mathrm{rank}(\bm{C}_k)=S]=}\nonumber\\
&=&
\P[\bm{\theta}_k^{\star} \textnormal{ is unique and } \mathrm{rank}(\bm{C}(\bm{\theta}_k^{\star}))=S]\nonumber\\
&=&\sum_{\theta=1}^H
\P[\bm{\theta}_k^{\star}=\theta , \mathrm{rank}(\bm{C}(\theta))=S].
\label{eq:Crankineq}
\eeqa
We now show that, for all $\theta=1,2,\ldots,H$: 
\beq
\boxed{
\P[\mathrm{rank}(\bm{C}(\theta))=S]=1
}
\label{eq:Cthetafull}
\eeq
It is useful to visualize the matrix $\bm{C}(\theta)$ as follows:
\beq
{\footnotesize
\left[
\begin{array}{cccc}
\bm{d_{\theta 1}} - \bm{d}_{11} &\bm{d_{\theta 2}} - \bm{d}_{12}&\ldots&\bm{d_{\theta S}} - \bm{d}_{1S}\\
\\
\bm{d_{\theta 1}} - \bm{d}_{21} &\bm{d_{\theta 2}} - \bm{d}_{22}&\ldots&\bm{d_{\theta S}} - \bm{d}_{2S}\\
\vdots&\vdots&&\vdots\\
\bm{d_{\theta 1}} - \bm{d}_{(\theta-1) 1} &\bm{d_{\theta 2}} - \bm{d}_{(\theta-1)2}&\ldots&\bm{d_{\theta S}} - \bm{d}_{(\theta-1)S}\\
\\
0&0&\ldots&0\\
\\
\bm{d_{\theta 1}} - \bm{d}_{(\theta+1)1} &\bm{d_{\theta 2}} - \bm{d}_{(\theta+1)2}&\ldots&\bm{d_{\theta S}} - \bm{d}_{(\theta+1)S}\\
\vdots&\vdots&&\vdots\\
\bm{d_{\theta 1}} - \bm{d}_{H 1} &\bm{d_{\theta 2}} - \bm{d}_{H 2}&\ldots&\bm{d_{\theta S}} - \bm{d}_{HS}\\
\\
1&1&\ldots&1
\end{array}
\right].
}
\label{eq:Bdelmat}
\eeq
The matrix $\bm{C}(\theta)$ has $H-1$ {\em random} rows (i.e., excluding the all-zeros and all-ones rows). 
Thus, when $H>S$ there are at least $S$ rows with random entries.
%%%%%
These random entries are jointly absolutely continuous since $i)$ so are the entries of $\bm{D}$; and $ii)$ the mapping from $\bm{D}$ to (the random entries of) $\bm{C}(\theta)$ is non-singular.\footnote{For example, property $ii)$ can be grasped by noting that, conditioned on $d_{\theta 1}, \ldots,d_{\theta S}$, the random entries in~(\ref{eq:Bdelmat}) are jointly absolutely continuous.} 
%%%%%
%It is straightforward to show that these random entries are distributed according to an absolutely continuous probability measure, 
This implies that, for $H>S$:
\beq
\P[\mathrm{rank}(\bm{C}(\theta))=S]=1,
\eeq
which proves~(\ref{eq:Cthetafull}) for the case $H>S$. 

We switch to the case $H=S$. 
Let us denote by $\bm{B}_{S-1}(\theta)$ the sub-matrix of $\bm{B}(\theta)$ obtained by deleting its last column, and with $\bm{b}_S(\theta)$ the last column of $\bm{B}(\theta)$. We can write:
\beq
\bm{C}(\theta)=\begin{bmatrix}
\bm{B}_{S-1}(\theta) &\bm{b}_S(\theta)\\
\mathbbm{1}_{S-1}\T&1 
\end{bmatrix}.
\label{eq:Busefulblock}
\eeq
We notice that $\bm{B}_{S-1}(\theta)$ depends only on the sub-matrix $\bm{D}_{S-1}$ that is obtained by deleting from $\bm{D}$ the last column. It is thus meaningful to introduce the set of matrices: 
\beq
\mathcal{E}\dfz\left\{D_{S-1}: \mathrm{rank}(B_{S-1}(\theta))=S-1\right\}.
\eeq
Recalling that $B_{S-1}(\theta)$ contains an all-zeros row, we see that, given a matrix $D_{S-1}\in\mathcal{E}$, there exists a unique sequence of weights:
\beq
w_1, w_2,\ldots,w_{\theta-1}, w_{\theta+1},\ldots,w_{S},
\label{eq:lincomb}
\eeq 
to obtain the row vector $\mathbbm{1}_{S-1}\T$ as a weighted linear combination of the rows of $B_{S-1}(\theta)$. 
Accordingly, given a matrix $D_{S-1}\in\mathcal{E}$, the rank of $\bm{C}(\theta)$ will be equal to $S$ if the last row in $\bm{C}(\theta)$ cannot be obtained as a linear combination of the rows of $\bm{B}(\theta)$. 
In view of~(\ref{eq:Busefulblock}), this corresponds to check whether the linear combination of the elements in $\bm{b}_S$  with the same weights is equal to $1$, namely, if:
\beq
\sum_{h\neq \theta} w_h (\bm{d}_{\theta S} - \bm{d}_{h S})=1. 
\label{eq:lincomblastcol}
\eeq
Consider now a matrix $D_{S-1}\in\mathcal{E}$ We have that:
\beq
\P\left.\left[\sum_{h\neq \theta} w_h (\bm{d}_{\theta S} - \bm{d}_{h S})=1 
\right|D_{S-1}
\right]=0,
\eeq
since (also conditioned on $\bm{D}_{S-1}$) the random variables $\{\bm{d}_{h S}\}$, with $h=1,2,\ldots,H$, are jointly absolutely continuous. We then conclude that: 
\beq
\P[\mathrm{rank}(\bm{C}(\theta))=S|D_{S-1}]=1,
\eeq
which implies~(\ref{eq:Cthetafull}) since, in view of the joint absolute continuity of the entries in $\bm{D}$, we have that: 
\beq
\P[\mathrm{rank}(\bm{B}_{S-1}(\theta))=S-1]=1 \Rightarrow \P[\bm{D}_{S-1}\in\mathcal{E}]=1.
\label{eq:rankBsm1}
\eeq
If we now apply~(\ref{eq:Cthetafull}) in~(\ref{eq:Crankineq}), we conclude that:
\beqa
\lefteqn{\P[\bm{\theta}_k^{\star} \textnormal{ is unique and }  \mathrm{rank}(\bm{C}_k)=S]=}\nonumber\\
&=&\sum_{\theta=1}^H \P[\bm{\theta}_k^{\star}=\theta]=\P[\bm{\theta}_k^{\star} \textnormal{ is unique }].
\eeqa
The proof of the theorem will be now complete if we show that the probability of having a unique $\bm{\theta}_k^{\star}$ is equal to $1$. To this aim, by using~(\ref{eq:uniquemin}) and~(\ref{eq:networkdivhomogen}), we see that:
\beq
\bm{\theta}_k^{\star}=\arg\!\min_{\theta\in\Theta}\sum_{s=1}^S x_{sk} \bm{d}_{\theta s}.
\label{eq:thetastarfinal}
\eeq
Let us consider the summations in~(\ref{eq:thetastarfinal}) corresponding to different values of $\theta$. 
Since the random variables $\{\bm{d}_{\theta s}\}$ are jointly absolutely continuous (and since $x_k$ is not an all-zeros vector), the probability that two or more summations are equal is zero, which finally implies that $\bm{\theta}_k^{\star}$ is unique.
~\hfill$\blacksquare$

\end{appendices}

%\bibliographystyle{IEEEtran}
%\bibliography{IEEEabrv,biblio,biblio2}

\begin{thebibliography}{10}
\providecommand{\url}[1]{#1}
\csname url@samestyle\endcsname
\providecommand{\newblock}{\relax}
\providecommand{\bibinfo}[2]{#2}
\providecommand{\BIBentrySTDinterwordspacing}{\spaceskip=0pt\relax}
\providecommand{\BIBentryALTinterwordstretchfactor}{4}
\providecommand{\BIBentryALTinterwordspacing}{\spaceskip=\fontdimen2\font plus
\BIBentryALTinterwordstretchfactor\fontdimen3\font minus
  \fontdimen4\font\relax}
\providecommand{\BIBforeignlanguage}[2]{{%
\expandafter\ifx\csname l@#1\endcsname\relax
\typeout{** WARNING: IEEEtran.bst: No hyphenation pattern has been}%
\typeout{** loaded for the language `#1'. Using the pattern for}%
\typeout{** the default language instead.}%
\else
\language=\csname l@#1\endcsname
\fi
#2}}
\providecommand{\BIBdecl}{\relax}
\BIBdecl

\bibitem{MattaSantosSayedICASSP2019}
V.~Matta, A.~Santos, and A.~H. Sayed, ``Exponential collapse of social beliefs over weakly-connected heterogeneous networks,'' in \emph{Proc. IEEE International Conference on Acoustics, Speech and Signal Processing (ICASSP)}, Brighton, UK, May 2019, pp. 5267--5271.

\bibitem{ChamleyBook}
C.~Chamley, \emph{Rational Herds: Economic Models of Social Learning}.\hskip
  1em plus 0.5em minus 0.4em\relax Cambridge, UK: Cambridge Univ. Press, 2004.

\bibitem{AcemogluOzdaglar2011}
D.~Acemoglu and A.~Ozdaglar, ``Opinion dynamics and learning in social networks,'' \emph{Dyn. Games Appl.}, vol.~1, no.~1, pp. 3--49, 2011.

\bibitem{PoorSPmag2013}
V.~Krishnamurthy and H.~V. Poor, ``Social learning and {B}ayesian games in
  multiagent signal processing: {H}ow do local and global decision makers
  interact?'' \emph{{IEEE} Signal Process. Mag.}, vol.~30, no.~3, pp. 43--57,
  May 2013.

\bibitem{Jadbabaie2013}
A.~Jadbabaie, P.~Molavi, and A.~Tahbaz-Salehi, ``Information heterogeneity and
  the speed of learning in social networks,'' \emph{Columbia Business School
  Research Paper}, pp. 13--28, May 2013.
    
\bibitem{ScaglioneSPmag2013}
C.~Chamley, A.~Scaglione, and L.~Li, ``Models for the diffusion of beliefs in
  social networks: An overview,'' \emph{{IEEE} Signal Process. Mag.}, vol.~30,
  no.~3, pp. 16--29, May 2013.

\bibitem{ScaglioneACM2013}
E.~Yildiz, A.~Ozdaglar, D.~Acemoglu, A.~Saberi, and A.~Scaglione, ``Binary
  opinion dynamics with stubborn agents,'' \emph{ACM Trans. Econ. Comput.},
  vol.~1, no.~4, pp. 19:1--19:30, Dec. 2013.

\bibitem{NedicTAC2015}
A.~Nedi\'{c} and A.~Olshevsky, ``Distributed optimization over time-varying
  directed graphs,'' \emph{{IEEE} Trans. Autom. Control}, vol.~60, no.~3, pp.
  601--615, Mar. 2015.

\bibitem{YingSayed2016}
B.~Ying and A.~H. Sayed, ``Information exchange and learning dynamics over
  weakly connected adaptive networks,'' \emph{{IEEE} Trans. Inf. Theory},
  vol.~62, no.~3, pp. 1396--1414, Mar. 2016.

\bibitem{Salami}
H.~Salami, B.~Ying, and A.~H. Sayed, ``Social learning over weakly connected
  graphs,'' \emph{{IEEE} Trans. Signal Inf. Process. Netw.}, vol.~3, no.~2, pp.
  222--238, Jun. 2017.

\bibitem{Zhao}
X.~Zhao and A.~H. Sayed, ``Learning over social networks via diffusion
  adaptation,'' in \emph{Proc. Asilomar Conference on Signals, Systems and
  Computers}, Nov. 2012, pp. 709--713.

%%%%%
%%%%%

\bibitem{SmithSorensen2000}
L.~Smith and P.~Sorensen, ``Pathological outcomes of observational learning,'' \emph{Econometrica}, vol.~68, no.~2, pp. 371--398, 2000.

\bibitem{Acemoglu2010}
D.~Acemoglu, M.~Dahleh, A.~Ozdaglar, and A.~Tahbaz-Salehi, ``Observational learning in an uncertain world,'' in \emph{Proc. IEEE Conf. Decision Control (CDC)}, Dec. 2010, pp. 6645--6650. 

\bibitem{Krishnamurthy2014}
V.~Krishnamurthy, O.~N. Gharehshiran, and M.~Hamdi, ``Interactive sensing and decision making in social networks,'' \emph{Found. Trends Signal Process.}, vol.~7, no.~1--2, pp. 1--196, Apr. 2014.
%%%%%
%%%%%


\bibitem{AcemogluRevEc2011}
D.~Acemoglu, M.~Dahleh, I.~Lobel, and A.~Ozdaglar, ``Bayesian learning in
  social networks,'' \emph{Rev. Econ. Studies}, vol.~78, no.~4, pp. 1201--1236, Oct. 2011.

\bibitem{DeGroot}
M.~H. DeGroot, ``Reaching a consensus,'' \emph{J. Amer. Statist. Assoc.},
  vol.~69, no. 345, pp. 118--121, 1974.

\bibitem{Epstein2010}
L.~G. Epstein, J.~Noor, and A.~Sandroni, ``Non-{B}ayesian learning,'' \emph{BE
  J. Theor. Econ.}, vol.~10, no.~1, pp. 1--20, 2010.

\bibitem{AcemogluGEB2010}
D.~Acemoglu, A.~Ozdaglar, and A.~ParandehGheibi, ``Spread of (mis)information
  in social networks,'' \emph{Games and Economic Behavior}, vol.~70, no.~2, pp.
  194--227, Nov. 2010.

\bibitem{Jad}
A.~Jadbabaie, P.~Molavi, A.~Sandroni, and A.~Tahbaz-Salehi, ``Non-{B}ayesian
  social learning,'' \emph{Games and Economic Behavior}, vol.~76, no.~1, pp.
  210--225, Sep. 2012. 

\bibitem{Jad2}
P.~Molavi, A.~Jadbabaie, K. R.~Rad, and A.~Tahbaz-Salehi, ``Reaching consensus with increasing information,'' 
\emph{{IEEE} J. Sel. Topics Signal Process.}, vol.~7, no.~2, pp. 358--369, Apr. 2013. 

\bibitem{NedicTAC2017}
A.~Nedi\'{c}, A.~Olshevsky, and C.~A. Uribe, ``Fast convergence rates for
  distributed non-{B}ayesian learning,'' \emph{{IEEE} Trans. Autom. Control},
  vol.~62, no.~11, pp. 5538--5553, Nov. 2017.

\bibitem{Javidi}
A.~Lalitha, T.~Javidi, and A.~D. Sarwate, ``Social learning and distributed
  hypothesis testing,'' \emph{{IEEE} Trans. Inf. Theory}, vol.~64, pp.
  6161--6179, Sep. 2018.

\bibitem{CT}
T.~Cover and J.~Thomas, \emph{Elements of Information Theory}.\hskip 1em plus
  0.5em minus 0.4em\relax John Wiley \& Sons, NY, 1991.

\bibitem{Sayed}
A.~H. Sayed, ``Adaptation, Learning, and Optimization over Networks,'' \emph{Found. Trends Mach. Learn.}, vol.~7, no. 4-5, pp. 311--801, 2014.

\bibitem{tomo}
V.~Matta and A.~H. Sayed, ``Consistent tomography under partial observations over adaptive networks,'' \emph{IEEE Trans. Inf. Theory}, vol.~65, no.~1, pp. 622--646, Jan. 2019.

\bibitem{SantosMattaSayedIT2019}
A.~Santos, V.~Matta, and A.~H. Sayed, ``Local tomography of large networks under the low-observability regime,'' \emph{IEEE Trans. Inf. Theory}, Oct. 2019, doi: 10.1109/TIT.2019.2945033.

\bibitem{mateos}
G.~Mateos, S.~Segarra, A.~Marques, and A.~Ribeiro, ``Connecting the dots: Identifying network structure via graph signal processing,'' \emph{IEEE Signal Process. Mag.}, vol.~36, no.~3, pp. 16--43, May 2019.

\bibitem{dokmanic2015euclidean}
I.~Dokmanic, R.~Parhizkar, J.~Ranieri, and M.~Vetterli, ``Euclidean distance
  matrices: Essential theory, algorithms, and applications,'' \emph{IEEE Signal Process. Mag.}, vol.~32, no.~6, pp. 12--30, Nov. 2015.

\bibitem{horn2012matrix}
R.~A. Horn and C.~R. Johnson, \emph{Matrix Analysis}.\hskip 1em plus 0.5em
  minus 0.4em\relax Cambridge University Press, 2012.

\bibitem{baksalary1979matrix}
J.~Baksalary and R.~Kala, ``The matrix equation $AX - YB= C$,'' \emph{Linear
  Algebra and its Applications}, vol.~25, pp. 41--43, Jun. 1979.

\bibitem{gower1985properties}
J.~C. Gower, ``Properties of Euclidean and non-Euclidean distance matrices,''
  \emph{Linear Algebra and its Applications}, vol.~67, pp. 81--97, Jun. 1985.

\end{thebibliography}


\end{document}